%% file: LCvoid_fin.tex
\documentclass[useAMS,usenatbib]{mn2e}
\usepackage{graphicx}
\usepackage{amssymb}
\usepackage{lscape}

\newcommand{\apj}{ApJ}

\newcommand{\aap}{A\&A}

\newcommand{\mnras}{MNRAS}

\newcommand{\kms}{km\,s$^{-1}$}

\newcommand{\HI}{H{\sc i}}
\newcommand{\sunn}{$_{\odot}$}
\newcounter{qub}
\newcommand{\qq}{\addtocounter{qub}{1}\arabic{qub}}

\DeclareRobustCommand{\ion}[2]{%
\relax\ifmmode
\ifx\testbx\f
{\mathrm{#1\,\textsc{#2}}}\else
{\mathrm{#1\,\mathsc{#2}}}\fi
\else\textup{#1\,{\mdseries\textsc{#2}}}%
\fi}

\title[Lynx-Cancer void galaxies. I. Sample]{Study of galaxies
in the Lynx-Cancer void. I.  Sample description}
\author[S. A. Pustilnik, A. L. Tepliakova]
{S. A. Pustilnik,$^{1,2}$\thanks{sap@sao.ru (SAP), arina@sao.ru (ALT)}
and A. L. Tepliakova$^1$\footnotemark[1] \\
$^1$ Special Astrophysical Observatory of RAS, Nizhnij Arkhyz,
Karachai-Circassia 369167, Russia\\
$^2$ Isaac Newton Institute of Chile, SAO branch, Nizhnij Arkhyz, 369167, Russia}

\begin{document}

\label{firstpage}

\date{Accepted April 2011 28; Received July, 2010 }

\pagerange{\pageref{firstpage}--\pageref{lastpage}} \pubyear{2011}

\maketitle

\begin{abstract}
The evolution of galaxies is influenced by the environment in which they
reside. This effect should be strongest for the least-mass and -luminosity
galaxies. To study dwarf galaxies in extremely low density environments we
have compiled a deep catalogue of dwarf galaxies in the nearby Lynx-Cancer
void. This void hosts some of the most metal-poor dwarfs known to date.
It borders the Local Volume at the negative supergalactic $Z (SGZ)$
coordinates and has the size of more than 16~Mpc. With a distance to its
centre of only 18 Mpc it is close enough to allow the search for the
faintest dwarfs. Within the void  75 dwarf (--11.9 $> M_{\rm B} > $ --18.0)
and 4 subluminous (--18.0 $ > M_{\rm B} > $ --18.4) galaxies have been
identified. We present the parameters of the void galaxies and give a
detailed analysis of the completeness of the catalogue as a function of
magnitude and surface brightness. The catalogue appears almost complete to
$M_{\rm B} <$ --14~mag, but misses part of the fainter low surface
brightness (LSB) face-on galaxies.
This sample of void galaxies builds the basis of forthcoming observational
studies that will give insight into the main stellar population,
\HI-mass-to-light ratio, metallicity and age for comparison with dwarfs in
higher
density regions. We briefly summarize the information on the unusual objects
in the void and conclude that their concentration hints that voids are
environments that are favourable for finding and studying unevolved dwarf
galaxies.
\end{abstract}


\section{INTRODUCTION}
\label{sec:intro}

The modern models of large-scale  structure and galaxy formation,
including the state-of-art N-body simulations, predict that galaxy
properties and evolution significantly depend on global environment
\citep[e.g.][and references therein]{Peebles01,MW02,Tully02,Gottlober03,
Hoeft06,Arkhipova07,Hahn07,Hahn09}.
While the effect of a denser environment on galaxy properties and evolution
has been known for a rather long time \citep[e.g.][]{HGC84,BG06}, the
role of the most rarefied environment such as voids on galaxy formation and
evolution is less studied, either theoretically and observationally.

The latter is due to observational selection effects.
Most galaxies with known radial velocities are found in
spectral surveys of magnitude-limited samples. Wide-field spectral
surveys have typical apparent magnitude limits corresponding roughly to
$B \sim$18~mag.
This apparent magnitude limit implies that for distances well beyond the
Local Supercluster ($cz > 5000-6000$~\kms), where the great majority of
large voids (with sizes of 20--40 Mpc) were found,
the faintest selected galaxies will have absolute magnitudes $M_{\rm B}$
of $\sim$--16~mag or brighter.
This implies that the study of distant voids is
limited to galaxies $\sim$3--4 mag fainter than $L^{*}$ galaxies
($M_{\rm B}^{*}\approx$--19.5 to --20.0~mag).
$L^{*}$ galaxies are typically the ones that mark the borders of voids.
Therefore, even the most advanced studies of the void galaxy population,
based
on very large samples with redshifts from the Sloan Digital Sky Survey
(SDSS)
\citep[e.g.][]{Sorrentino06,Patiri06} of $z <$ 0.03--0.05
($M_{r} \lesssim -20.0$), were limited to galaxies with luminosities of
only $\sim$2 magnitudes fainter than $M_{\rm r}^{*}$. Only studies of
photometric and spectroscopic properties for SDSS samples with $z <$ 0.025,
$M_{r} \lesssim -14.0$ \citep{Rojas04,Rojas05} were able to probe less
luminous galaxies. A new detailed study of the properties
of galaxies in voids, described by \citet{Stanonik09} and \citet{vandeWey09}
also mainly  deals with more luminous dwarf galaxies, namely with
$M_{\rm R} <$  --16.0~mag.

The smaller a galaxy, the more fragile it is in respect to external
disturbance. Therefore, the possible difference of galaxy properties in
various types of environments is expected to depend on galaxy mass. Thus,
the
relatively `shallow' probes of `distant' void galaxy population based on the
SDSS samples leave significant room for deeper insight into the
question.

% -----------------------------------------------
%  Table of Nearby Voids parameters
% -----------------------------------------------

\begin{table*}
\centering
\caption{Parameters of the nearby voids}
\label{tab:voids}
\begin{tabular}{lrrrrrrrrr} \\ \hline \hline
%Designation& Size & RA    & Dec  & $cz$ &l$^{II}$& b$^{II}$&SGX&SGY &SGZ  \\
Designation        & Size & RA    & Dec  & $cz$ & $l$    & $b$    &$SGX$&$SGY$ &$SGZ$  \\
		   & \kms & hour  & \degr& \kms & \degr  & \degr  &\kms&\kms&\kms \\ \hline
Cetus              & 500  & 02.0  & --20  & 700  & 192    &$-72$   &100 &--600&--200 \\
Cepheus            & 500  & 23.5  & +65  & 800  & 112     &  +05   &700 &   0& 300  \\
Crater             & 500  & 11.5  & --15  & 1500 & 126    &$-28$   &1300&--700& 200  \\
Volans             & 700  & 07.0  & --70  & 800  & 281    &$-25$   &--600&--300&--500 \\
Monoceros          &1000  & 08.0  & +05   & 800  & 216    & +17    & 200& 430&--970 \\
Lynx-Cancer main   & 1200  & 07.9  & +27  & 1030 & 194    & +25    & 660& 660&--930 \\
Lynx-Cancer subvoid& 870  & 08.5  & +29  & 770  & 195    & +34    & 470& 674&--680 \\
Inner Local Void   &2000 & 18.5 & --01 & 900& 30    & +02    &--500&--200& 700 \\ \hline
\end{tabular}
\end{table*}

Earlier studies found galaxies in voids to typically be low-mass
actively star-forming galaxies, like BCGs and H{\sc ii} galaxies
\citep[in particular,][]{Salzer89,Pustilnik95,Popescu97}.
\citet{Lindner06} have shown that BCGs are not the only objects in
voids.  More typical dwarfs with lower star formation (SF) activity also
populate voids.
Statistical studies of optical properties of void galaxies
in the SDSS data did not provide further clues to the evolutionary state of
these galaxies, but confirmed the increased fraction of blue galaxies and
Star Formation Rate (SFR).
The only proxy of evolutionary parameters for void galaxies studied up
to now, is the ratio $M(HI)/L_{\rm B}$, but only for a limited
sample of BCGs
\citep{Pustilnik02}. The recent study of void galaxies via \HI\ imaging by
\citet{Stanonik09} is less selective, but still is rather limited with
regard to luminosity.

The amount of evidence that  BCGs in underdense regions  may represent
a less evolved population was growing during the last decade
\citep[e.g.][]{Peebles01,Pustilnik03,HS0837,HS2134}. However, this
might be due to selection effects which favour actively star-forming
galaxies. Therefore, there is a need to address this issue
by defining samples that also contain the more typical late-type galaxies,
particularly because void environments are thought to be favourable for
sufficiently quiet galaxy evolution. Hence, one can hope to use the
statistics of void galaxy ensembles as a good instrument for comparison with
predictions of cosmological galaxy formation scenarios.
Such large and deep samples of void galaxies would also allow a more
detailed comparison with predictions from cosmological N-body simulations
\citep[e.g.][]{Tikhonov09}.

To construct void galaxy samples with absolute magnitudes down to
$M_{\rm B}\sim$--12, from  samples with the typical apparent magnitude
limit of $m_{\rm B} \sim$18--19, one needs to limit the distant boundary by
$D \sim$10--16~Mpc. The nearest
voids \citep[e.g.][]{Fairall98}, adjacent  to the Local Volume
\citep[distances of $<$10~Mpc, as defined, e.g. by][]{CNG} are suitable
for this task.
Due to their relative proximity  ($D_{\rm centre} \sim$10--15 Mpc), the
low-luminosity galaxies can be relatively easily identified. Moreover,
spectroscopic studies of their element abundances in the relatively faint
H{\sc ii} regions of these fairly nearby galaxies is feasible
with modern large telescopes.

In this paper we present the sample of galaxies residing in one of the
nearest, Lynx-Cancer, voids mentioned in
\citet{Pustilnik03}. The choice of the void is motivated by a good coverage
of this sky region by the SDSS spectral and image databases
\citep{SDSS_phot,Gunn98,York2000,SDSS_phot1,Pier03}, which leads to
a significant increase in the number of known void galaxies and provides
the photometric properties of the void galaxies. The study of this sample
is also motivated by the prominent concentration of atypical dwarf galaxies
in this volume. Half a dozen objects
in this relatively modest void are very metal-poor galaxies and/or
reveal no visible old stellar population
\citep[][Pustilnik et al. 2011, MNRAS, submitted]{Pustilnik03,DDO68,J0926}.

The lay-out of the paper is as follows. In Section~\ref{sec:void_desc} we
summarize information on the nearby voids,  and describe the
Lynx-Cancer void. Section \ref{sec:void_dwarfs} presents the void galaxy
sample. In Section~\ref{sec:dis} we discuss the completeness of the void
galaxy sample, summarize the properties of the most unusual void galaxies
and the prospects to increase this galaxy sample with LSB dwarf galaxies
that are missed in the SDSS spectral database. Section~\ref{sec:summ}
presents the summary of the main results.
We adopt the value of Hubble constant as $H_{0}$=73~\kms~Mpc$^{-1}$.

\section{Nearby voids. The Lynx-Cancer void description}
\label{sec:void_desc}

\subsection{Nearby voids}
\label{sec:nearbyvoids}

\begin{figure*}
 \centering
 \includegraphics[angle=-0,width=8cm]{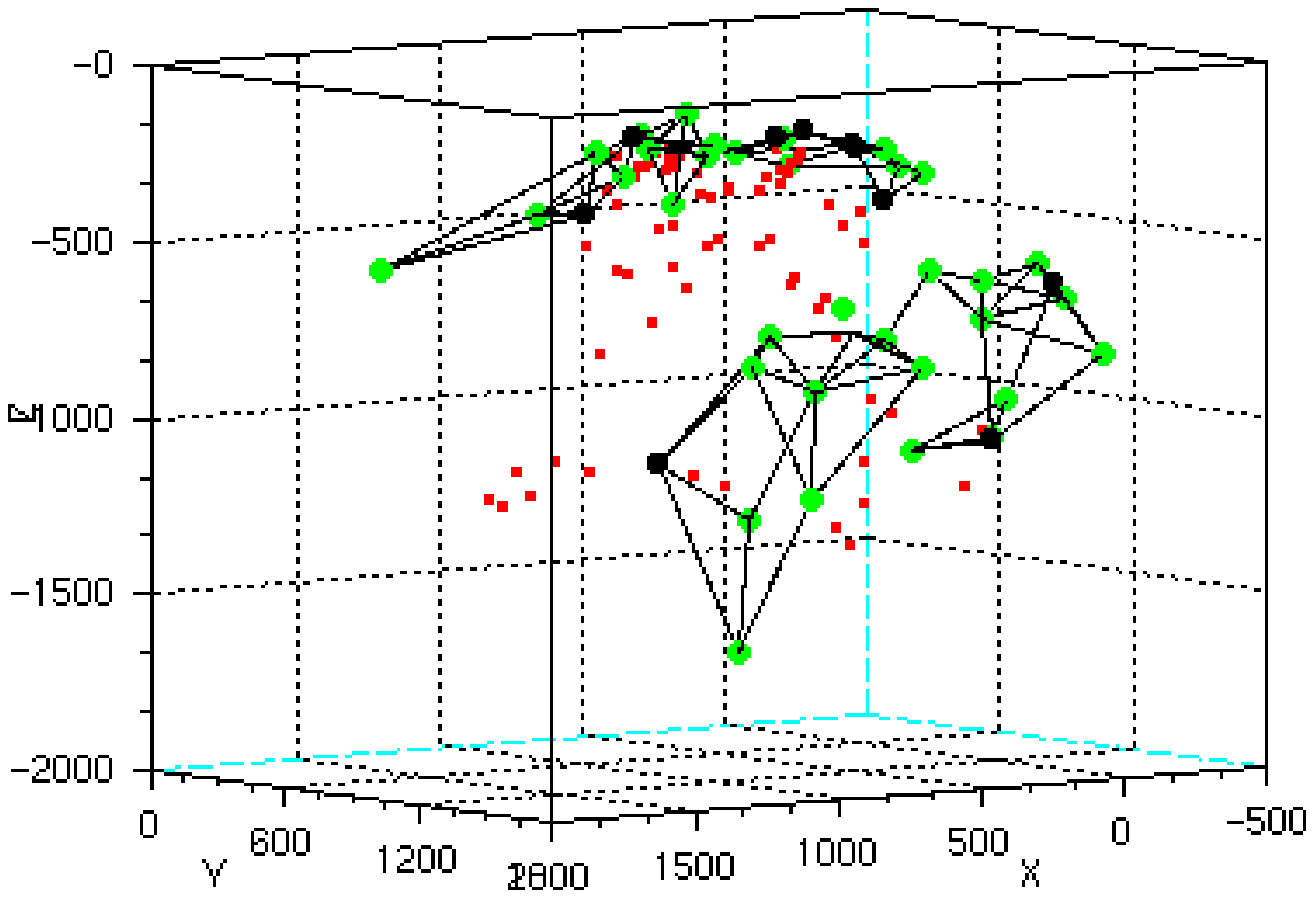}
 \includegraphics[angle=-0,width=8cm]{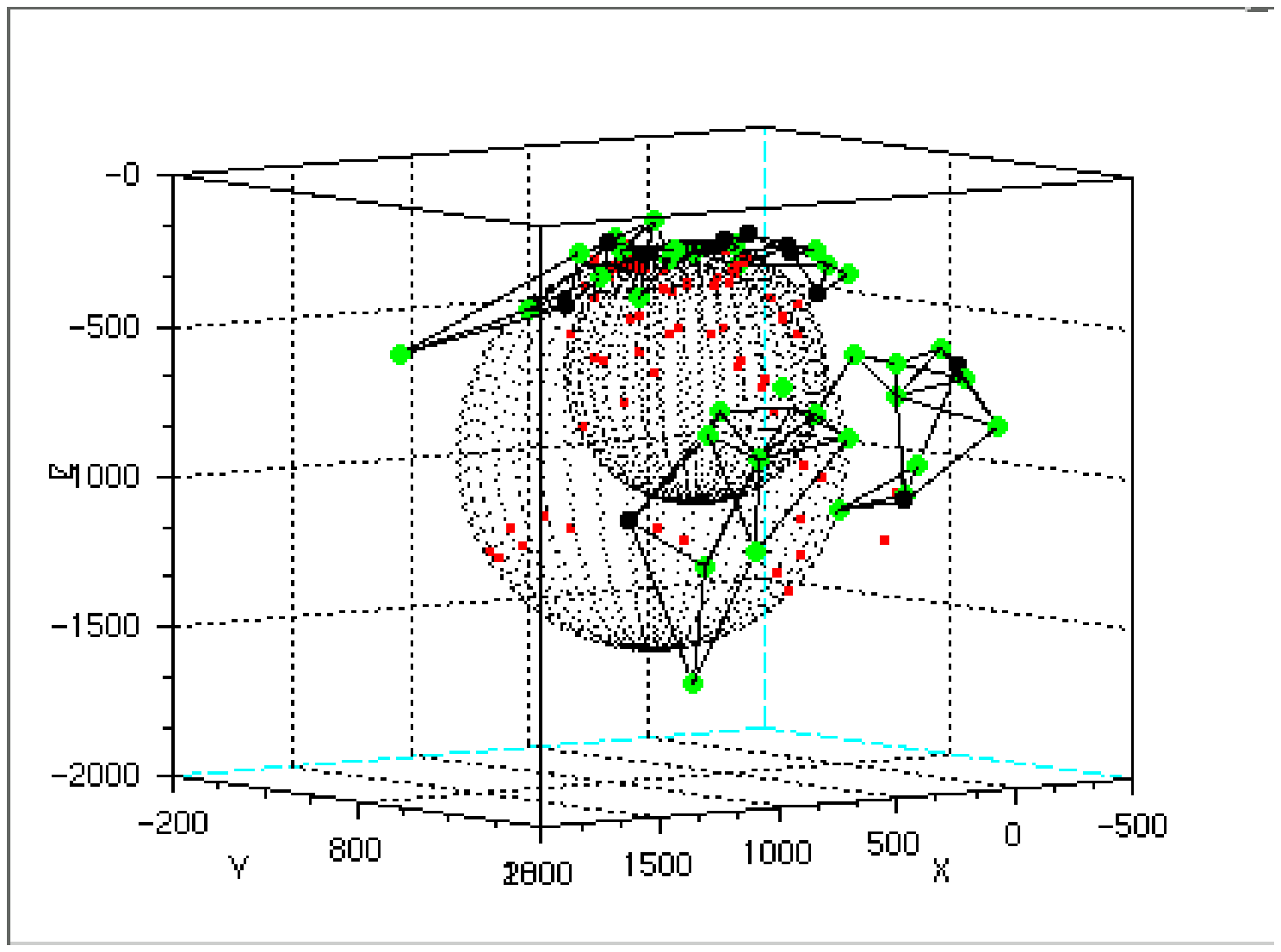}
  \caption{\label{fig:void_3d}
{\it Left panel:} the 3D view of the Lynx-Cancer void region with positions
of `luminous' galaxies ($M_{\rm B} < $--19.0~mag). The supergalactic
coordinate
grid ($SGX, SGY, SGZ$) is in units of \kms. Large filled green dots show
positions of pairs and groups in which at least one `luminous' galaxy enters.
Large black filled dots show positions of `isolated' luminous galaxies. Each
luminous galaxy is joined with the three the nearest luminous neighbours to
help visually define the void border. Dwarf galaxies inside the void are
shown by small red squares.
{\it Right panel:} The same figure with a contour sketch of the larger main
void (large sphere) and the subvoid (smaller sphere), suggested in
\citet{Pustilnik03}, which includes a large fraction of void galaxies.
}
\end{figure*}

The issue of the nearest voids was briefly addressed by
\citet{Fairall98}. In Table~\ref{tab:voids} we summarise (ranked on the
void size, apart from the Lynx-Cancer subvoid) and update on the main
parameters of the nearby voids. In comparison to the
original \citet{Fairall98} list, the Inner Local Void from \citet{Tully08}
is added. The misprint is corrected for RA of the Monoceros void centre
(A.~Fairall, private communication) and for all the related parameters as
well. Finally, the Lynx-Cancer void and its subvoid are added with their
parameters. The void maximal extent in units of \kms\
is shown  in column 2. In columns 3 and 4 we present the approximate
equatorial coordinates of the void centres.
Column 5 gives the distances to  the void centres ($cz$
relative to the restframe of the Local Group in \kms).  In columns 6 and 7
the Galactic longitudes and latitudes of the void centres are given,
respectively, while columns 8, 9 and 10 present their approximate
supergalactic $X, Y$ and $Z$ coordinates, respectively.

The giant Local Void described by \citet{Tully08} begins close to the Local
Group and its neighbouring groups at the positive supergalactic $Z$.
The Local Void appears pretty empty. Its expected population
of dwarf galaxies remains almost elusive. Whether this is the
effect of obscuration by the Galaxy disc or the intrinsic property of the
Local Void, or the combination of both, remains unclear.
While other nearby voids are also interesting for detailed studies of their
galaxy population, we concentrate here on the Lynx-Cancer void.

The Lynx-Cancer void was discovered as a result of a study of the very
low-metallicity BCG HS~0822+3542 \citep[one of many such galaxies found in
the Hamburg--SAO Survey;][and references therein]{HSS1,HSS6} and its
companion, the LSB dwarf galaxy SAO 0822+3545. \citet{Pustilnik03}
noticed  that they are situated in a very rarefied environment. This was
the motivation for a more careful analysis of the galaxy content in this
volume  which led to the identification of another nearby void,
similar to those described by \citet{Fairall98}. This void is adjacent to
the Monoceros void and they probably represent the parts of a larger
'empty' volume at the negative $SGZ$. In comparison to other nearby voids,
the Lynx-Cancer void has two advantages:
(a) the major part of the respective sky region is covered by the SDSS
imaging and spectroscopy database
\citep[][and references therein]{York2000,DR7}, and (b) a large part of this
sky region will be covered by the blind high-sensitivity HI surveys,
including the Arecibo survey ALFALFA \citep[e.g.][]{ALFALFA}, the Australian
survey ASKAP \citep{ASKAP}, the Westerbork survey Apertif
\citep[e.g.][]{Apertif} and the Effelsberg survey EBHIS
\citep[e.g.][]{EBHIS}.

Surprisingly, several other very metal-poor dwarf galaxies, including  two
objects  with $Z \lesssim$  1/30~$Z$\sunn\
[or 12+$\log(O/H) \lesssim$7.16], were found in this volume
\citep[][see details in Section~\ref{sec:dis}]{DDO68,IT07,J0926}.
These and other findings hint at the probable
effect of the void environment on dwarf galaxy formation and evolution.
However, to study the evolutionary status of void dwarf galaxies
statistically, one needs to describe the void geometry and boundaries
more carefully, to  better  define the dwarf galaxy sample falling
within the void. The statistical approach also requires a well-defined
control sample in regions of higher galaxy densities. Fortunately, the
detailed studies of a large number of dwarf galaxies have appeared recently,
which can be used for comparison; such as the FIGGS sample of 65 faint
($M_{\rm B} \gtrsim $ --15.7) late-type dwarfs
\citep{Begum08}, of which $\sim$3/4 belong to the Local Volume groups.

\subsection{Luminous galaxies delineating the Lynx-Cancer void}
\label{sec:borders}

We show the boundaries of the Lynx-Cancer void, as delineated by
individual `luminous' galaxies (here, with $M_{\rm B} < -$19.0),
or the individual galaxies of  pairs and groups (large dots) in
Fig.~\ref{fig:void_3d}.  A sample of luminous galaxies, delineating the
void, is presented and discussed  below. Their lists are given in Tables
\ref{tab:lum} and \ref{tab:groups} in Appendix~\ref{App1}.

Constructing this bordering luminous/massive object sample, we first picked
up a greater number of luminous objects in the surrounding volume, defined
by the sky region with R.A. between 6~h and 11~h and Declinations of
$>$ 0\degr\ and radial velocities $V_{\rm hel} < $ 1800~\kms, or distances
$D \lesssim$ 30 Mpc.
Then, analysing the intermediate results in 3D pictures, we reduced the
sample of luminous galaxies in order to leave only those objects which
really define the boundaries of this void. For galaxies with  known
non-redshift distances, either from  the photometric methods - cepheid,
Tip of Red Giant Branch  (TRGB), or surface-brightness fluctuations (SBF)
or from the Tully-Fisher luminosity-rotation velocity relation \citep[see
`The Extragalactic Distance Database',][]{Tully09}, we computed
the `distance' velocity as follows: $V_{\rm dist}$ = 73~$\times$~$D$(Mpc).

\citet{Tully08} noted that galaxies in the region considered here
have a large peculiar radial velocity component.  This is induced by
the giant Local Void being situated approximately on the opposite side
of the Local Sheet. The geometry is illustrated in Fig.~\ref{fig:slices}
in the appendix. It shows on a larger scale the position of the
Lynx-Cancer void in supergalactic coordinates in two slices, together
with both the Local Sheet and the Local Void.  The Local Sheet galaxies
(including the Local Group), being at the Local Void border, move with
$V$=323~\kms\ along the vector {\bf \it n}, directed to the sky position
with coordinates of $l$=220\degr, $b$=32\degr\ \citep{Tully08}.
To account for this large peculiar velocity, the galaxies for which
only redshifts are available require a correction
$V_{\rm dist} = V_{\rm LG} + V_{\rm pec}$,
where $V_{\rm pec}$ = 323$\times$($\cos\theta$)~\kms\ with
$\theta$ being the angle between the vector {\bf \it n} and direction
to a galaxy.

To verify the choice of such a correction, we compared the resulting
$V_{\rm dist}=V_{\rm LG}+V_{\rm pec}$ for 21 galaxies in the considered
region, for which the independent reliable distances are known from TRGB,
cepheid or SBF methods. The derived weighted mean difference
of --9$\pm$32~\kms\ is consistent with no systematic difference between
the real value and that derived through $V_{\rm pec}$.

\subsection{The Lynx-Cancer void description}
\label{sec:LCdescript}

Real voids are not perfect spheres. But as the first
approximation of a void, we use the largest empty spheres which can be
inscribed in the distribution of dots,
representing `luminous' galaxies.  To find such spheres, we accounted for
the borders of the studied volume mentioned in Sec.~\ref{sec:borders}.
The largest sphere appeared significantly larger and more distant
(`Lynx-Cancer main' void in Table 1) than that previously suggested
in \citet{Pustilnik03}. However, we also identified a smaller sphere,
significantly intersecting with the former one (called the Lynx-Cancer
subvoid). Its parameters are similar to those of the originally described
Lynx-Cancer void from \citet{Pustilnik03}. Both of these spheres are shown
in contours in the right-hand panel of Fig.~\ref{fig:void_3d}. The centre
of the main void with radius $R=$8.2~Mpc is at distance $D=$18.0~Mpc. The
centre of the subvoid with $R=$6.0~Mpc is at $D=$14.6~Mpc.

As described above, in many cases the form of real voids is nonspherical.
Also, the
modelling of voids  \citep[e.g.][and references therein]{Lavaux10}
indicates that smaller voids are expected to be more elongated. Therefore,
for the subsequent analysis we not only assign to the Lynx-Cancer void
region the interior of the two maximal spheres described above, which
includes 45 galaxies with $D_{\rm NN} >$ 2~Mpc, but also
examine galaxies in the adjacent empty regions between the bordering
luminous objects and the surfaces of these empty spheres. If these
galaxies fall in the regions situated far from luminous galaxies
($D  > $ 2 Mpc),
these regions are treated also as parts of the void and the related
galaxies are also  qualified as void objects.
About half of the dwarf galaxies classified as void objects in
Section~\ref{sec:void_dwarfs}  belong to these empty regions outside the
well-defined maximal void spheres.

\section{The dwarf galaxy sample within the Lynx-Cancer void}
\label{sec:void_dwarfs}

A sample of 75 dwarf ( --11.9 $> M_{\rm B} >$--17.9) and 4 subluminous
[$M_{\rm B} \sim$--(18.0--18.3)] galaxies, falling within the Lynx-Cancer
void, is presented in Table \ref{tab:voidgal}.
To separate this sample, we proceeded as follows. First, we
selected in two steps all isolated galaxies within the void region. At the
first step, all galaxies were considered isolated if they had a projected
distance to a luminous nearest neighbour $>$ 1 Mpc. At the second step, for
the void galaxies with  luminous neighbours closer than 1 Mpc in
projection,
we used a finer criterion, corresponding to the results on satellites
of massive galaxies, presented by \citet{Prada03}.

In the analysis of a very large sample of SDSS galaxies, \citet{Prada03}
have shown that for a galaxy with luminosity $L^{*}$, the
relative r.m.s. line-of-sight velocities of satellites, $\sigma_{\rm vel}$,
change from 120~\kms\ at 20~kpc to $\sim$80~\kms\ at 200~kpc, and to
 $\sim$60~\kms\ at 350~kpc. The satellite velocities at large distances
scale as $L^{0.5}$ of the host galaxy. To qualify the status of a small
galaxy, we took these results into account. Namely,
based on the $M_{\rm B}$ and the projected distance of the nearest luminous
galaxy, we estimated the respective value of 2$\sigma_{\rm vel}$
following the results of \citet{Prada03}.
If $| \Delta$V$_{\rm rad}|$ for the small galaxy in  question is larger
than the  estimated 2$\sigma_{\rm vel}$, this dwarf was treated
as unrelated to the luminous neighbour, that is as an isolated object.
Following \citet{Kara_groups05}, we adopted orbital masses in the range
(2.6$\pm$1.3)$\times$10$^{12}$~$M$\sunn\ for groups. According to
\citet{Prada03}, this implies values of $\sigma_{\rm vel}$ of 65-100~\kms\
at $R=$350~kpc, and a factor of  1.5 larger at $R=$100~kpc.
Once  the isolated galaxies are selected, the distance to the
nearest luminous neighbour $D_{\rm NN}$ was then calculated as
the length  of the 3D radius-vector between the two objects.

Finally, we assign an isolated galaxy to the void galaxy sample
if it falls inside the maximal sphere of the main Lynx-Cancer void or
its smaller subvoid, as described above, and have $D_{\rm NN}$ $>$
2.0 Mpc.  As noted in Section~\ref{sec:LCdescript},
we also included in the void sample all isolated galaxies which, due
to non-sphericity of the void, appear in the adjacent regions somewhat
outside the respective spheres  and  have $D_{\rm NN}$ $>$ 2.0 Mpc.

For two late-type spiral galaxies IC~2233 and NGC~2537, situated on the sky
close to each other, we used distances from the detailed study by
\citet{MU08}, which finds no evidence or traces of interaction between
these two galaxies.
A good TRGB distance estimate is known only for the almost edge-on LSB
spiral IC~2233. We adopt for this galaxy the mean distance module of
three discussed values by \citet{MU08}: $\mu_{\rm mean}$=30.15 mag,
respectively  10.7$\pm$0.5~Mpc. The less accurate
NGC~2537 distance estimators are consistent with the galaxies to be
unrelated \citep[see][for references and detailed discussion]{MU08}.
We adopted for the latter galaxy the distance determined from its radial
velocity and general peculiar velocity correction, which appears consistent
with the estimates in literature  within their accuracies.

In Table \ref{tab:voidgal} we present the following information on galaxies
falling into the Lynx-Cancer void. \\
Column 1. Common name or SDSS prefix. \\
Columns 2 and 3. Epoch J2000 R.A. and Declination. \\
Columns 4 and 5. $V_{\rm hel}$ and its error (almost all are either from NED
or SDSS). For several objects either without SDSS/NED velocities, or for
which this is significantly improved or corrected, we give the latter
values.
They include PGC2807187 \citep{KMKM08}; pair HS~0822+3542 and SAO~0822+3545,
KISSB~23 and SDSS~J0926+3343 \citep{Chengalur06,PM07,J0926}; UGC~3912
\citep{Springob05}, SDSS~J0723+3621, SDSS~J0723+3622, SDSS~J0737+4724,
SDSS~J0852+1351 (Pustilnik et al., MNRAS, submitted),
% MCG9-13-56 (SDSS DR7),
NGC~2537 and IC~2233 \citep{MU08}. \\
Column 6. The respective $V_{\rm LG}$. \\
Column 7. The velocity $V_{\rm dist}$ (in \kms) corresponds to
  $D$(Mpc)$\times$73 when the photometric (TRGB, cepheids) distance estimate
$D$(Mpc) is available (only for 4 galaxies, marked by $*$, typical accuracy
$\sim$10\%). Otherwise, this is $V_{\rm LG}$ + $V_{\rm pec}$, where
$V_{\rm pec} \sim$300~\kms\ is a correction for the peculiar velocity
described in Section~\ref{sec:borders}.
In this case the distance accuracy comes from the quadratic sum of
$\sigma(V_{\rm hel}$) and $\sigma(V_{\rm pec}$). The first term
is $<$10--15~\kms\ for about 3/4 of all sample galaxies, rising on average
to $\sim$30--40~\kms\ for the rest objects. The second term is 25~\kms\
\citep{Tully08}. Hence, for about 3/4 of the sample galaxies the typical
uncertainty $\sigma(V_{\rm dist}$) is $<$30~\kms, or
$\sigma(D)\lesssim$0.4~Mpc. For the remaining galaxies, the typical
$\sigma(V_{\rm dist}$) is $\sim$40--45~\kms, or $\sigma(D)\sim$0.6~Mpc. \\
Columns 8 and 9. $B_{\rm tot}$ (from NED or from the literature) and the
Galaxy extinction $A_{\rm B}$ \citep[from NED, following][]{Schlegel98}.
The sources of $B$-mag are given by a letter in the superscript for the
respective values in the following order. $b$, \citet{KMKM08}; $c$,
\citet{LEDA00,LEDA03}; $d$, \citet{Barazza01}; $e$, \citet{RC3.9}; $f$,
mean of photographic magnitudes from UGC and MCG catalogs, transformed to
$B$-band; $g$, \cite{KKH01}; $h$, \citet{CNG}; $i$, \citet{Zee00}; $j$,
\cite{Garnier96}; $l$, \citet{MU08}; $m$, \citet{Pustilnik03}; $n$,
\citet{J0926}; $o$, \citet{Salzer02}; $p$, \citet{DDO68}, $q$,
Pustilnik et al., MNRAS, submitted. When $B_{\rm tot}$ was unavailable
in the literature, we used the brightest (model) values of the SDSS $g$ and
$r$-filter magnitudes of those presented in NED (if present, or directly
from SDSS DR7) and transformed them to $B$-band magnitudes according to the
\citet{Lupton05}
formulae [$B = g + 0.3130(g - r) + 0.2271$;  $\sigma$ = 0.0107].
This case is marked by superscript $k$ after the value of $B_{\rm tot}$. The
resulting rms accuracy $\sigma_{\rm B}$ for the great majority of the sample
galaxies is $\lesssim$0.1--0.2~mag. However, for several galaxies outside
the SDSS zone, for which the CCD photometry appeared unavailable, the
$\sigma_{\rm B}$ can be as large as $\sim$0.5~mag. \\
Column 10. It presents $M_{\rm B}$, corrected for $A_{\rm B}$, calculated
from values in Columns 8 and 9 and the distance $D$(Mpc), corresponding to
$V_{\rm dist}$, that is $D$=73$V_{\rm dist}$. \\
Column 11. (Tentative) morphological class. \\
Column 12. The distance in Mpc to the nearest luminous galaxy or group,
$D_{\rm NN}$. \\
Column 13. Either an alternative name or some important comments, like
the presence of companion, etc.

\begin{figure*}
 \centering
 \includegraphics[angle=-0,width=16cm]{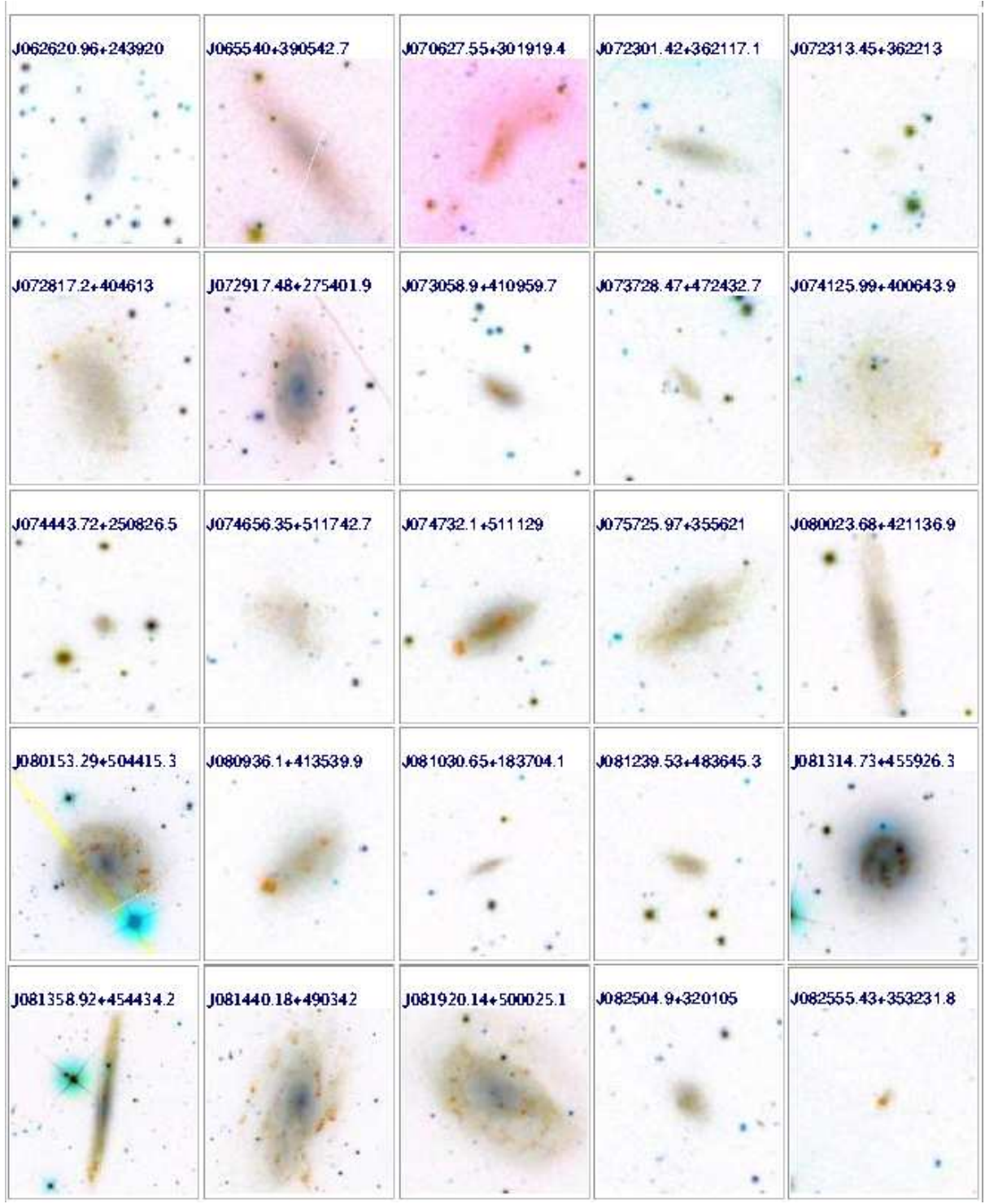}
  \caption{\label{fig:void_fc1}
The finding charts of 25 galaxies from the Lynx-Cancer void sample
(see Table~\ref{tab:voidgal}), falling in the zone covered by the SDSS.
They are prepared with the SDSS DR7 Navigate Tool with inverted colours and
are shown in the order of their RA. The side of each square measures
$\approx$100\arcsec, with except of several larger galaxies. For J0729+2754,
J0801+5044, J0813+4559 the side is $\approx$200\arcsec, while for
J0813+4544, J0814+4903, J0819+5000 the side is $\approx$300\arcsec.
}
\end{figure*}

\begin{figure*}
 \centering
 \includegraphics[angle=-0,width=16cm]{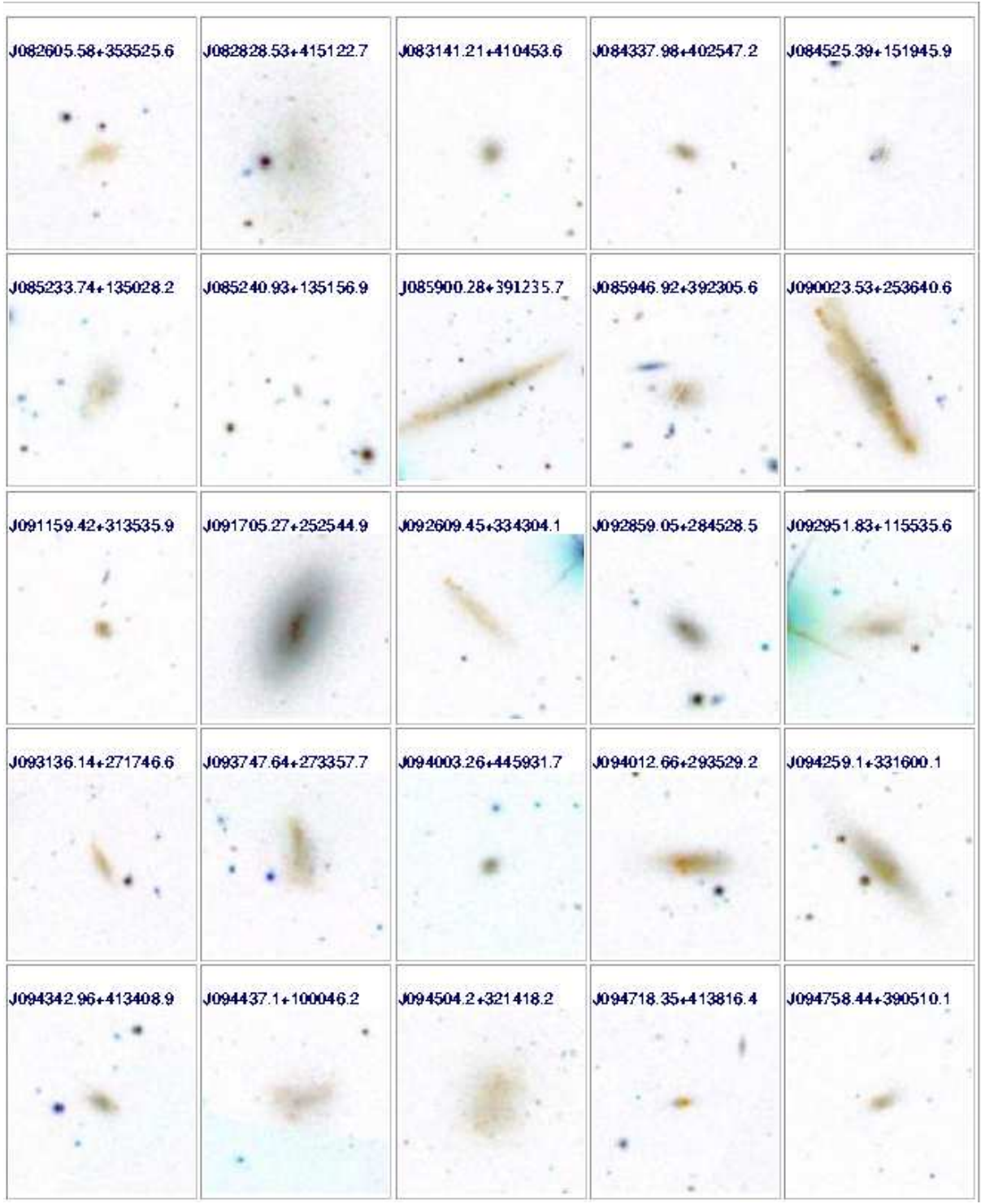}
  \caption{\label{fig:void_fc2}
The same as for Fig.~\ref{fig:void_fc1}, for the next 25  galaxies from
the Lynx-Cancer void sample, falling in the zone covered by the SDSS. The
finding chart size for a larger galaxy J0859+3912 is $\approx$200\arcsec.
}
\end{figure*}

\begin{figure*}
 \centering
 \includegraphics[angle=-0,width=16cm]{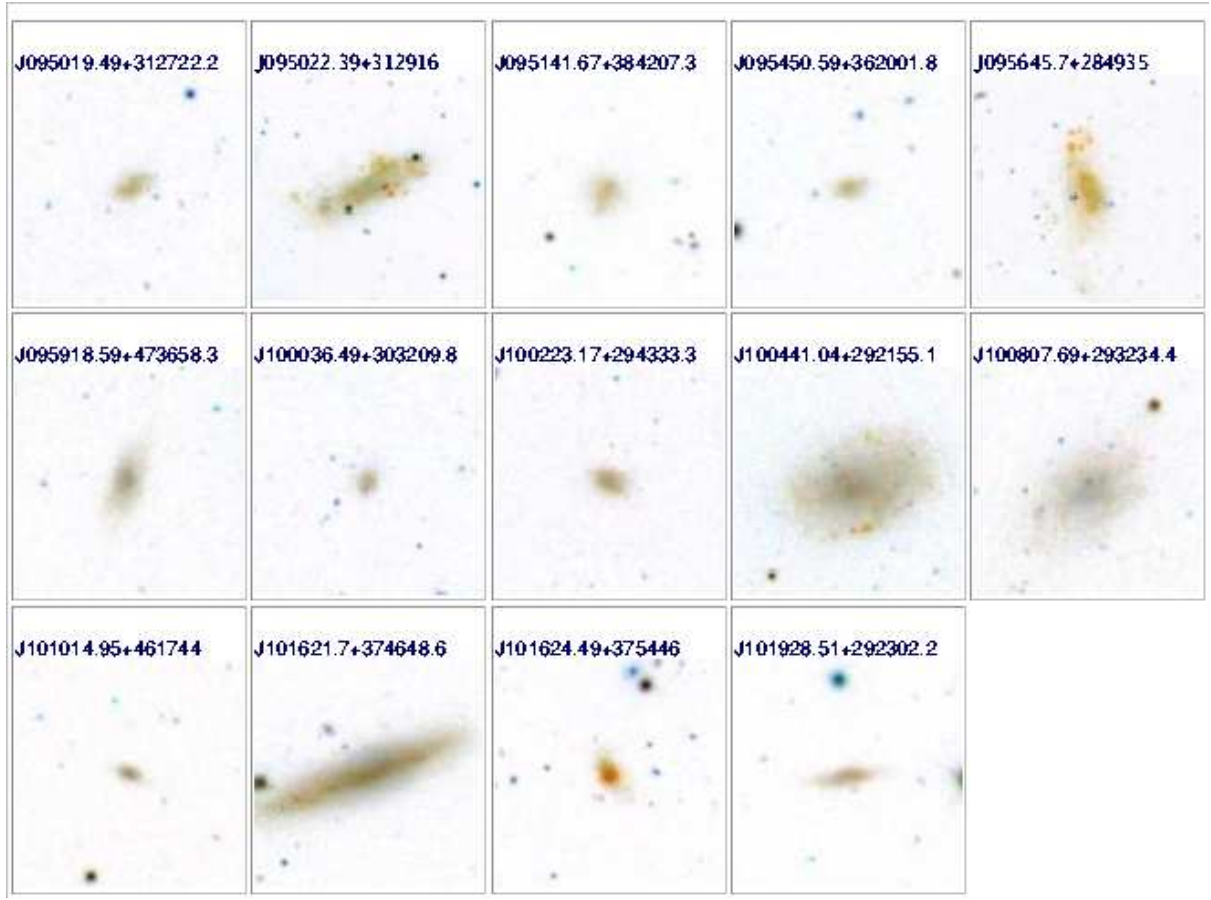}
  \caption{\label{fig:void_fc3}
The same as for Fig.~\ref{fig:void_fc1}, for the remaining 14  galaxies from
the Lynx-Cancer void sample, falling in the zone covered by the SDSS.
The finding chart size of two larger galaxies (J0950+3129, J0956+2849)
is $\approx$200\arcsec.
}
\end{figure*}

\begin{figure*}
% \centering
\centerline{\includegraphics[angle=-0,width=3.2cm]{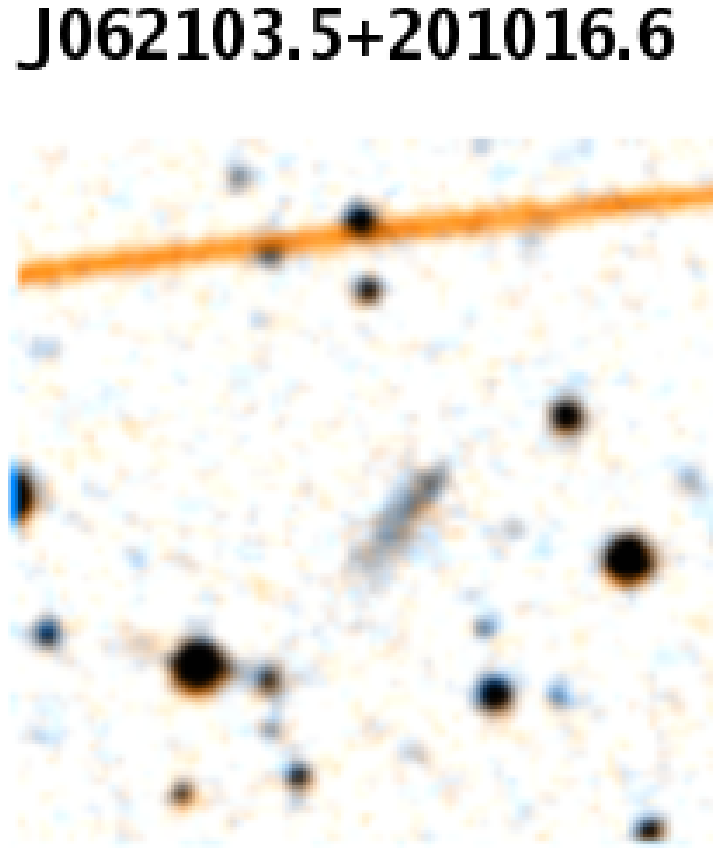}
 \includegraphics[angle=-0,width=3.2cm]{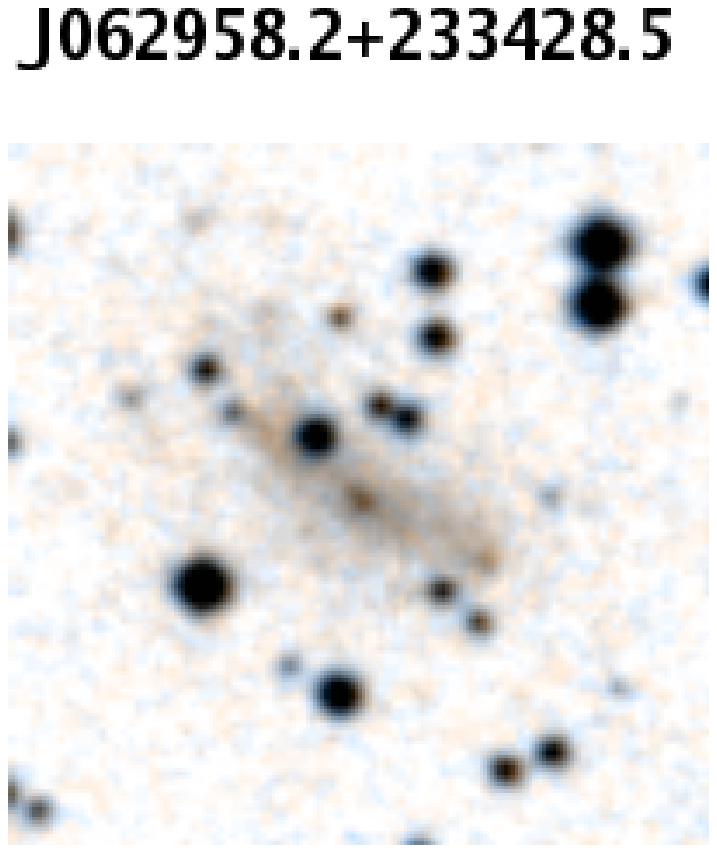}
 \includegraphics[angle=-0,width=3.2cm]{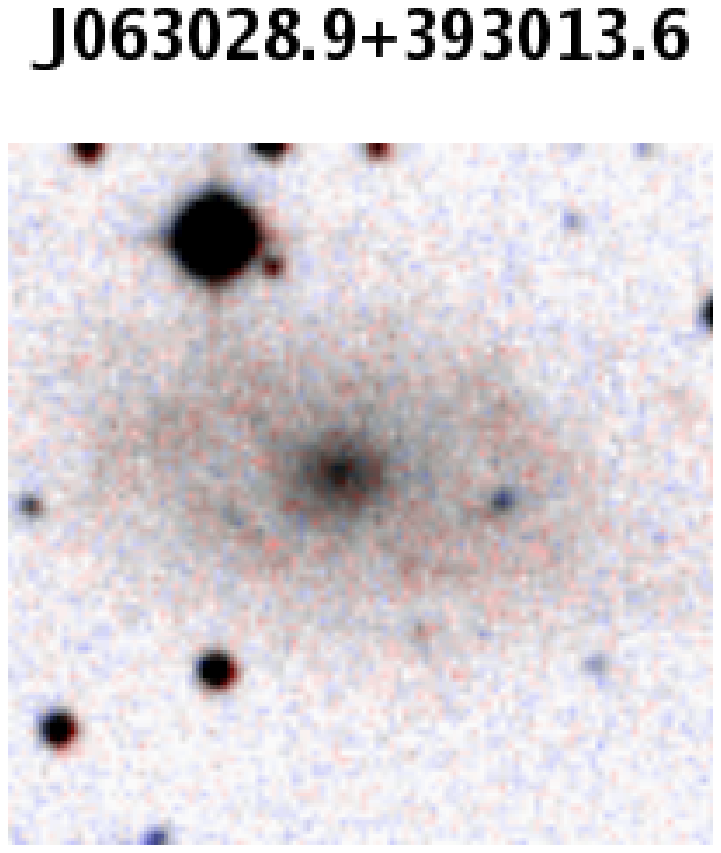}
 \includegraphics[angle=-0,width=3.2cm]{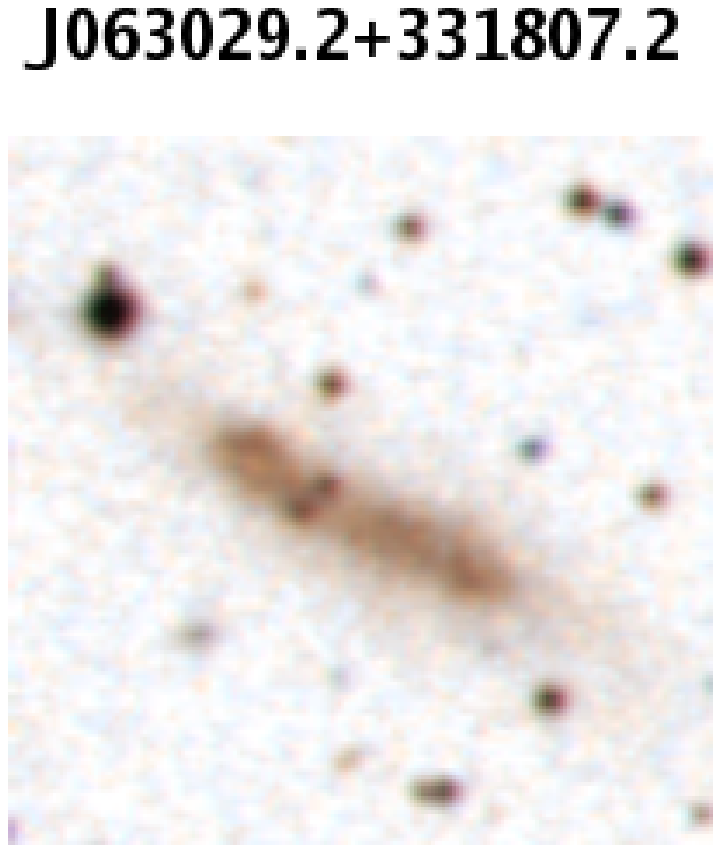}
 \includegraphics[angle=-0,width=3.2cm]{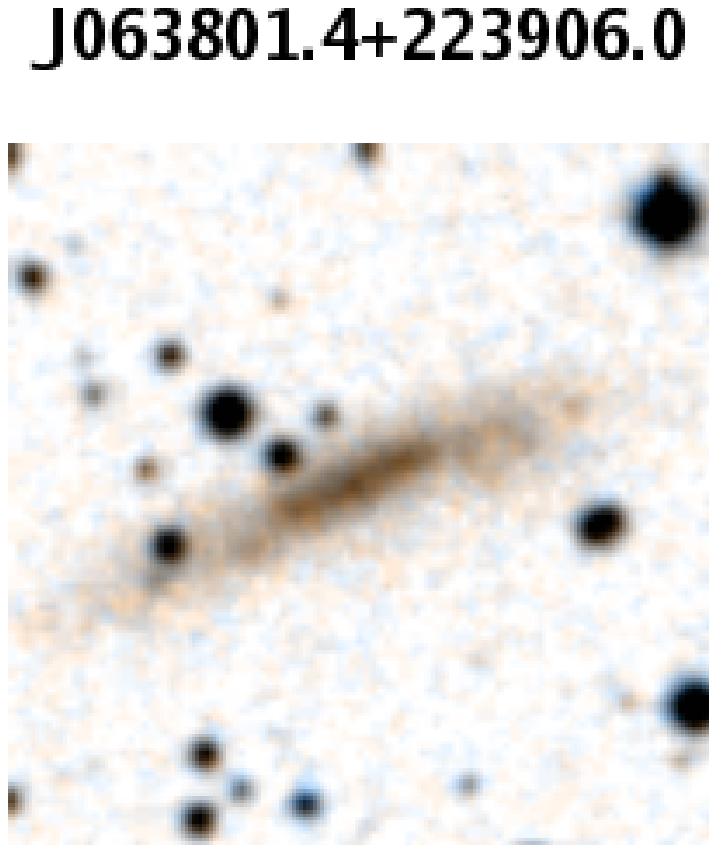}}
\centerline{\includegraphics[angle=-0,width=3.2cm]{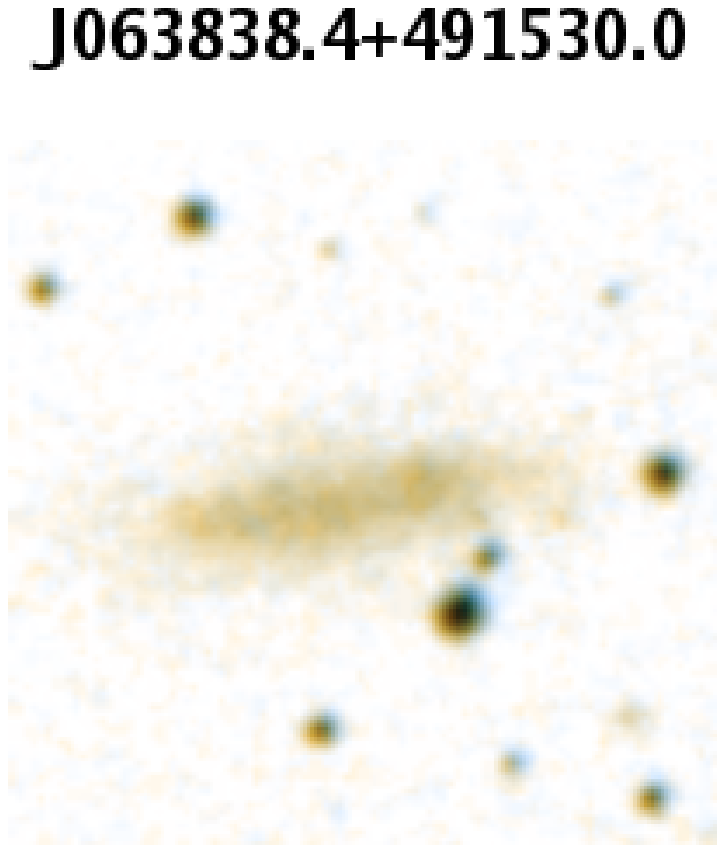}
 \includegraphics[angle=-0,width=3.2cm]{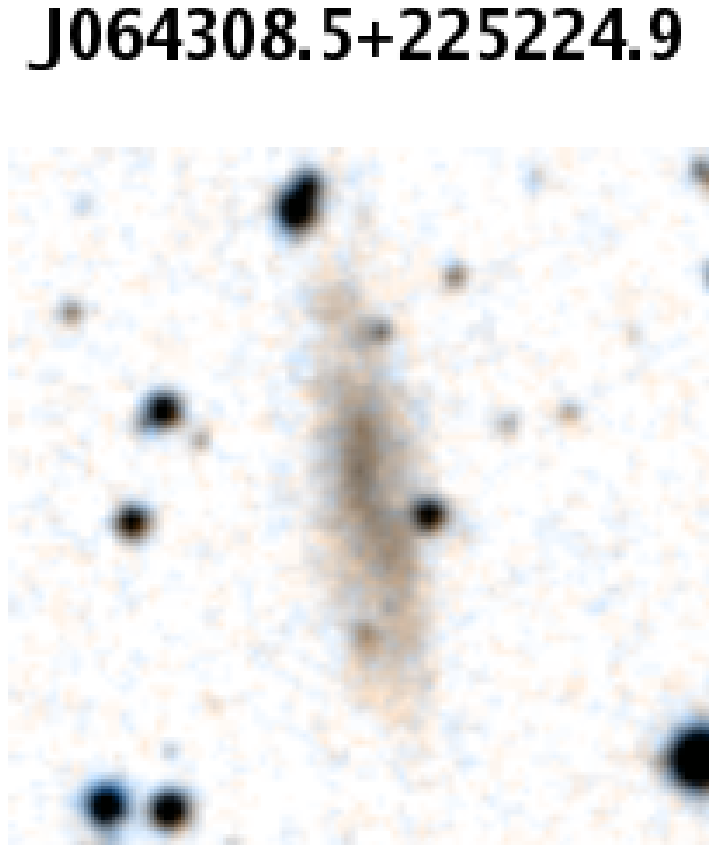}
 \includegraphics[angle=-0,width=3.2cm]{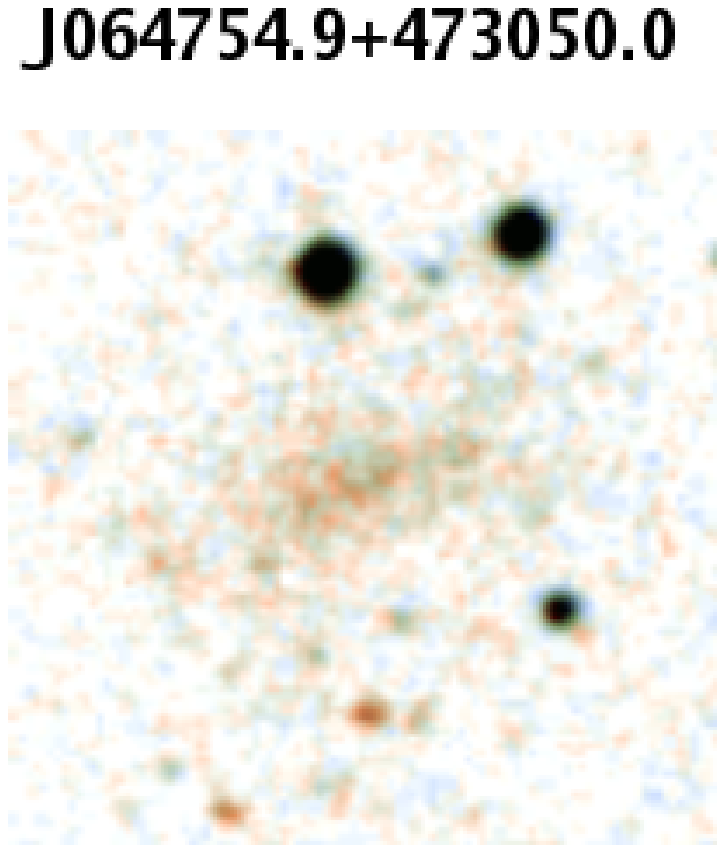}
 \includegraphics[angle=-0,width=3.2cm]{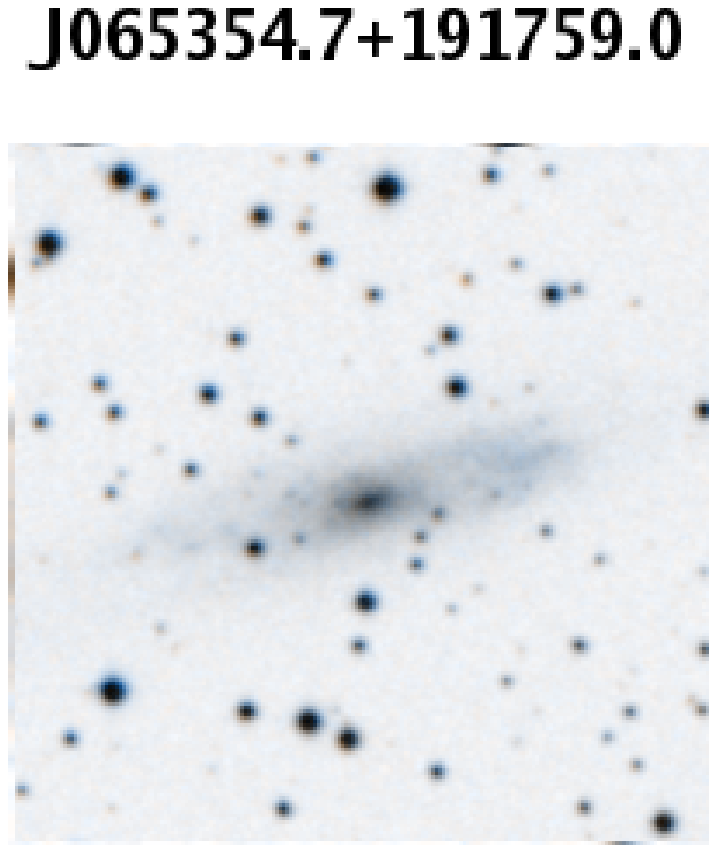}
 \includegraphics[angle=-0,width=3.2cm]{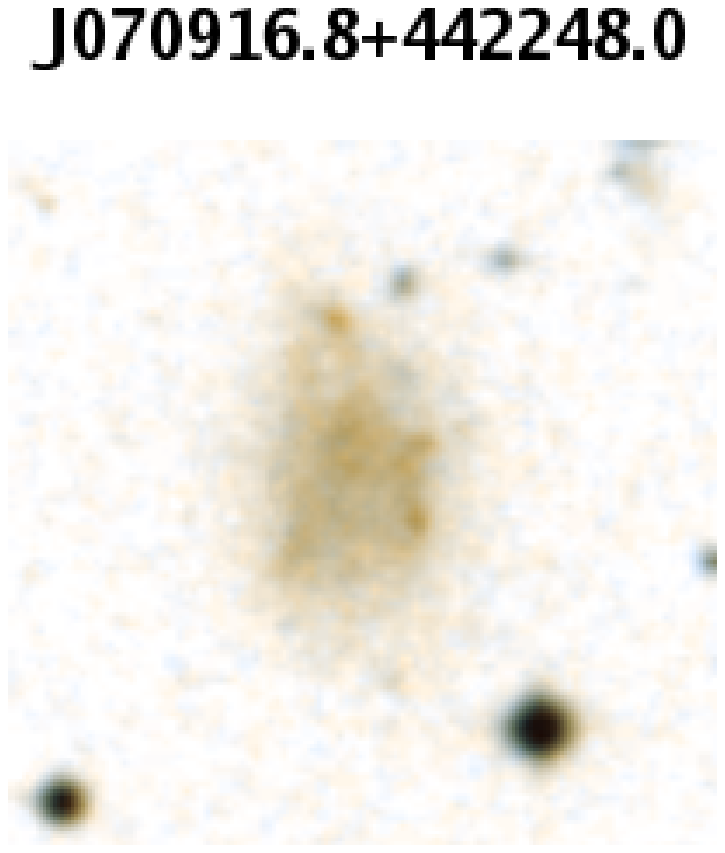} }
\centerline{\includegraphics[angle=-0,width=3.2cm]{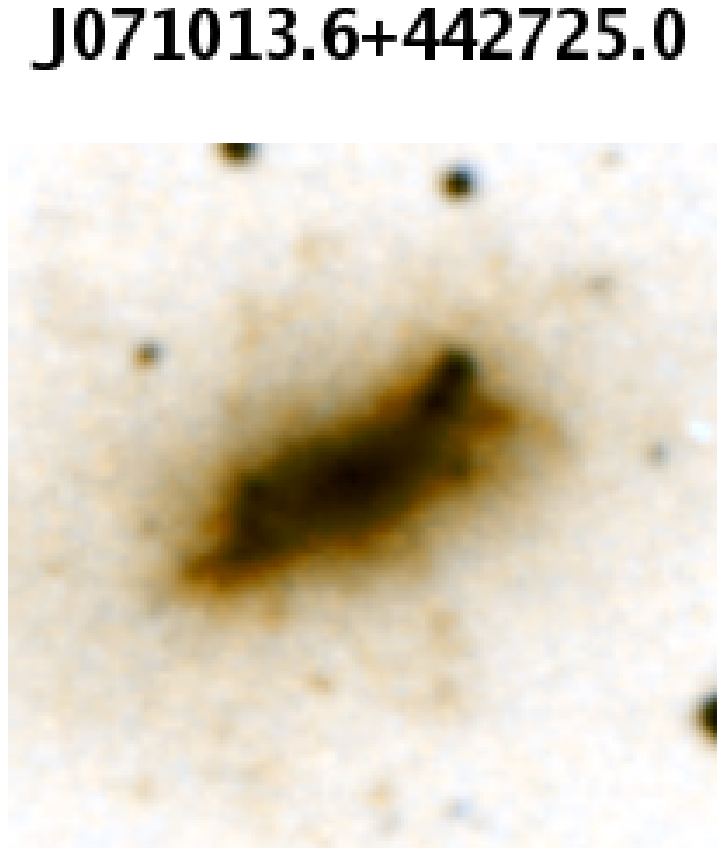}
 \includegraphics[angle=-0,width=3.2cm]{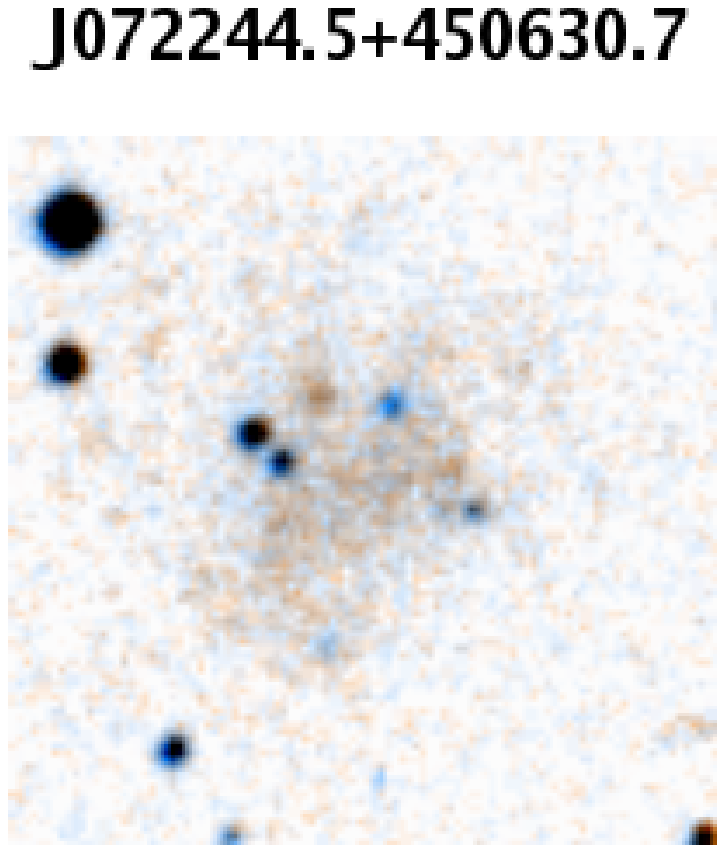}
 \includegraphics[angle=-0,width=3.2cm]{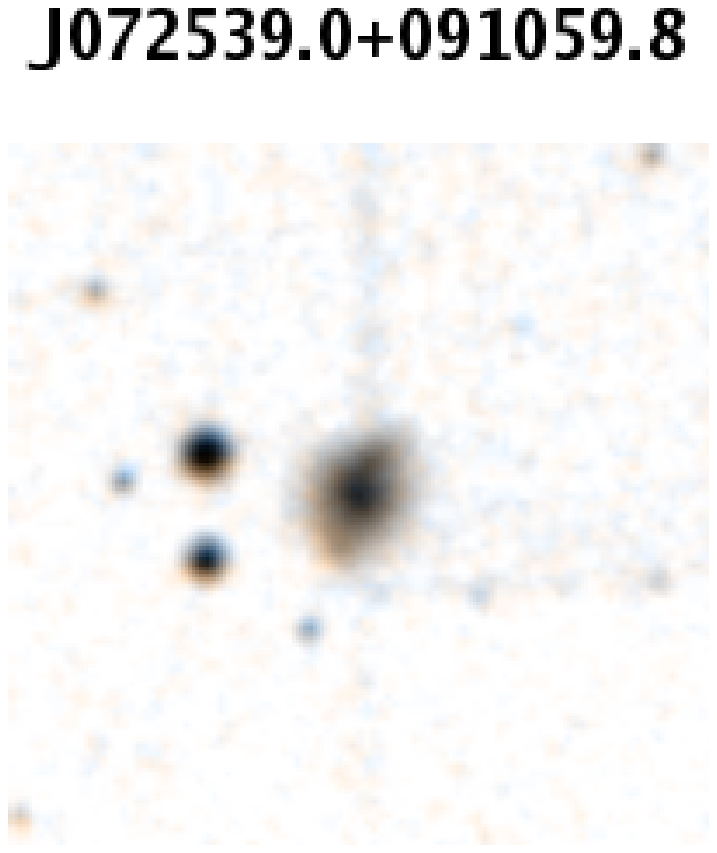}
 \includegraphics[angle=-0,width=3.2cm]{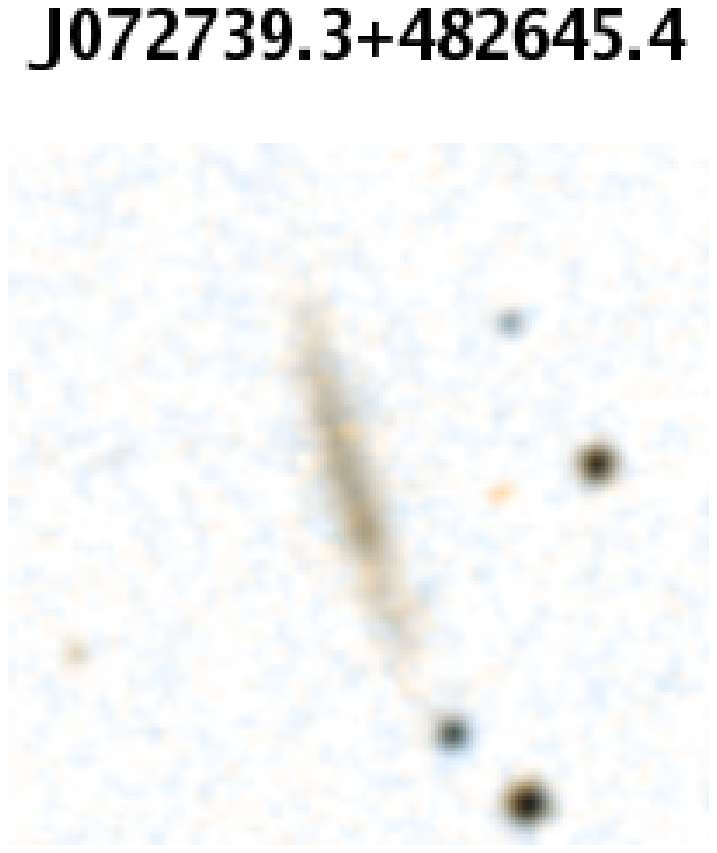}
 \includegraphics[angle=-0,width=3.2cm]{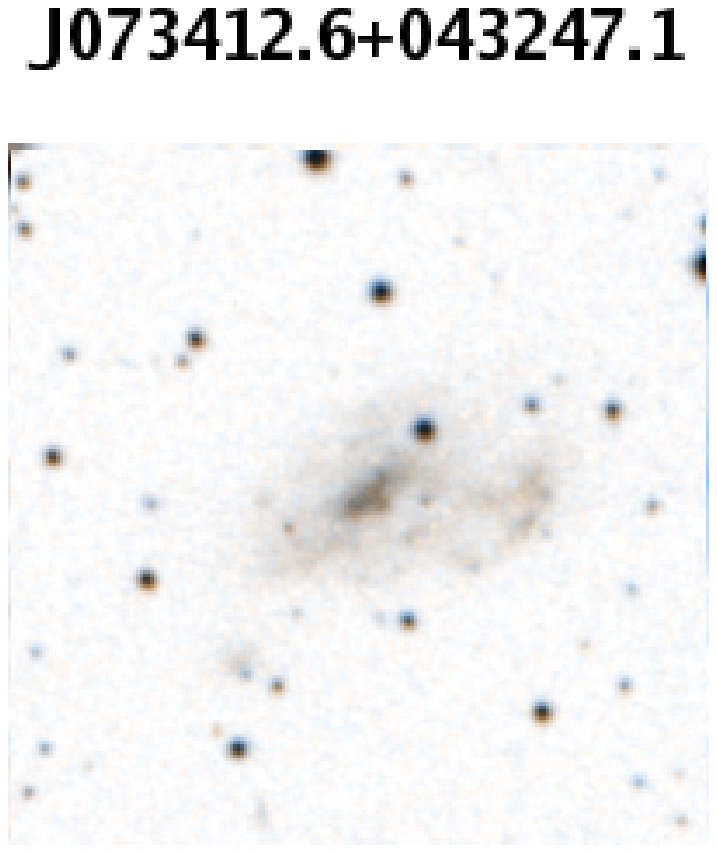}}
  \caption{\label{fig:void_fc4}
Same as in Fig.~\ref{fig:void_fc1}, but for 15 galaxies from the Lynx-Cancer
void sample falling outside the SDSS footprint. They are prepared
combining  the $O$ and $E$ images of the Digitized Palomar Sky Survey
and are shown in order of their Right Ascension. The colour balance is
very approximate. For larger galaxies J0653+1917 and J0743+0432 the size of
the chart is $\approx$200\arcsec. }
\end{figure*}

In Figs~\ref{fig:void_fc1} - \ref{fig:void_fc3} we present the mosaic of
64 void galaxy images prepared with the SDSS Navigate Tool.
For the remaining 15 void galaxies situated outside the SDSS footprint,
we show their images in Fig.~\ref{fig:void_fc4},
as extracted from the combined Palomar Sky Survey $O$ and $R$ images.

\section{Properties of the void galaxy sample and related issues}
\label{sec:dis}

\subsection{Global properties}

\begin{figure*}
 \centering
 \includegraphics[angle=-90,width=7.5cm]{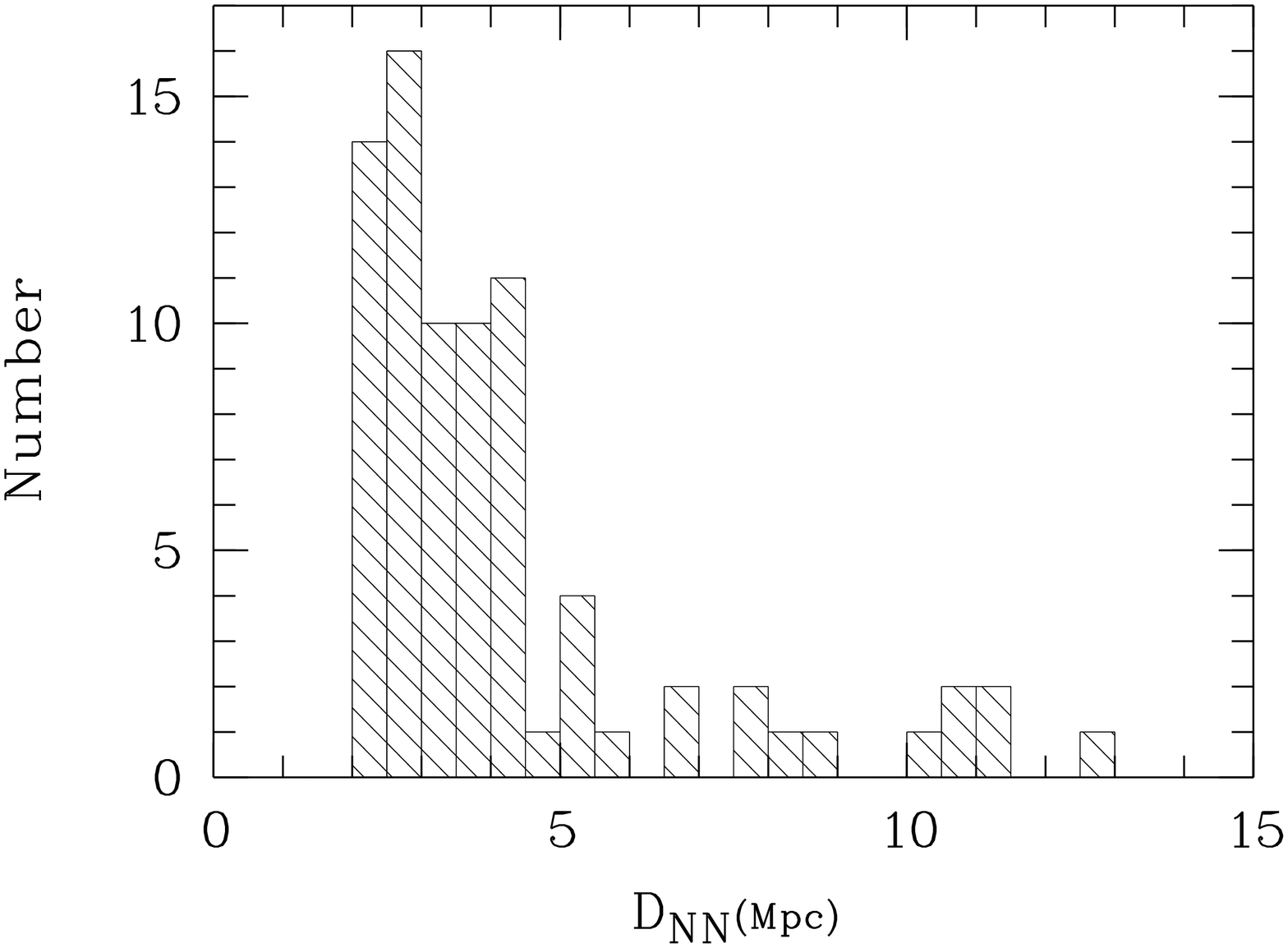}
 \includegraphics[angle=-90,width=7.5cm]{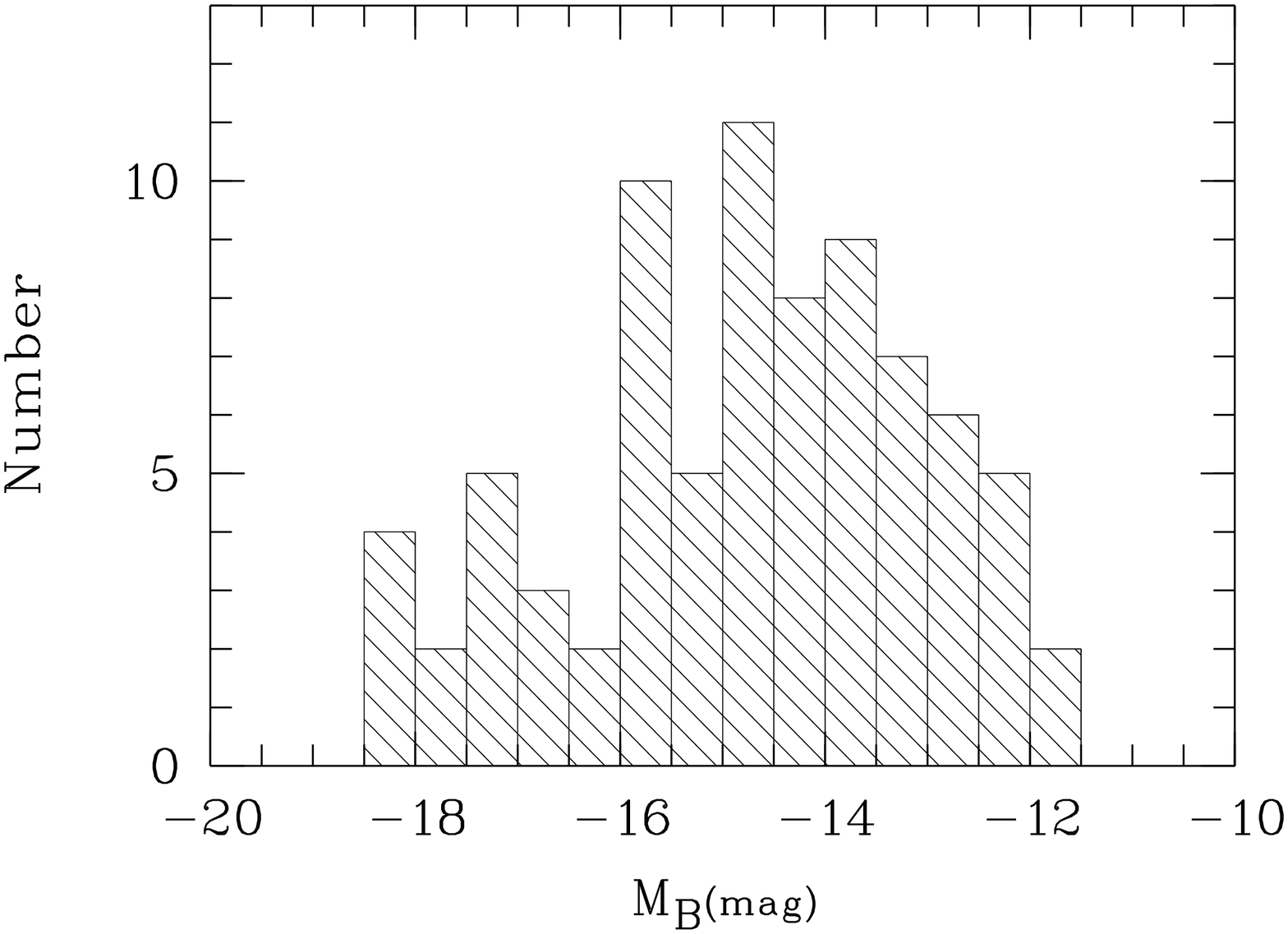}
  \caption{\label{fig:histograms}
{\it left panel:} the distribution of distances to the nearest luminous
galaxy $D_{\rm NN}$ for  the void galaxy sample.
{\it right panel:} the
distribution of absolute blue magnitudes $M_{\rm B}$ for  the
void galaxies.}
\end{figure*}

The main goal of selecting a galaxy sample falling inside the Lynx-Cancer
void is to study evolutionary parameters and to perform a comparison
with similar parameters of dwarfs in a more typical and denser environment.
The evolutionary parameters can also depend on galaxy global parameters,
such as mass or luminosity. Therefore, we  present and briefly summarize the
distributions of void  galaxies  with respect to $D_{\rm NN}$, $M_{\rm B}$
and morphological types. As one can see in the left-hand panel of
Fig.~\ref{fig:histograms}, about half of the void galaxies
have $D_{\rm NN}$ in the range of 2.0--3.5~Mpc, which is close to the
adopted threshold value of 2.0~Mpc. This corresponds to the known fact
that a large fraction of void galaxies reside close to the void
boundaries. Very isolated galaxies, with $D_{\rm NN}$ from 5 to
13~Mpc comprise $\sim$16\% of the sample. The blue
luminosities of the void galaxies, $M_{\rm B}$, spread over a range of
more than six magnitudes (Fig.~\ref{fig:histograms}, right-hand panel),
from $\sim$--12.0 to --18.3, with the median and peak value of the
histogram near $\sim$--14.5. The latter
indicates the substantial incompleteness of this void sample for
$M_{\rm B} > $  --14.0.

A preliminary visual analysis of  the void galaxy images in
Figs~\ref{fig:void_fc1}--\ref{fig:void_fc4} shows that about half of them
appear Low
Surface Brightness (LSB) objects. Two objects are classified as blue compact
galaxies (BCGs). These are HS~0822+3542 and HS 1013+3809, both having very
low O/H. They both are the components of pairs with more massive galaxies
\citep{Kniazev98,Kniazev00,Pustilnik03,Kniazev03,PM07}. One more galaxy -
NGC~2537=Mkn~86 is also classified as a BCG in some works. The remaining
objects are either more or less typical late-type dwarf spirals or irregular
galaxies with the `normal' level of current SF. More accurate data on
the LSB galaxy fraction will be given in a forthcoming paper where the
fitting surface brightness (SB) radial profiles is performed and the value
of central SB of the underlying discs is estimated.

\subsection{Completeness of the void galaxy sample}

The study of  the luminosity function of void galaxies for absolute
magnitudes $M_{\rm B} \le$ --14 is very important for comparison with the
results of CDM cosmological simulations. In particular, the question of how
faint and slow-rotating galaxies are found within voids was recently
addressed by \citet{Tikhonov09}. The latter issue requires a careful
analysis
of selection effects, which limit the number of known faint galaxies in the
void sample. A better understanding of the spatial structures (filaments)
in the void galaxy distribution can also be used as a test of structure
formation in low-density regions \citep{ParkLee09}.

It is well established that the luminosity and central SB of galaxies are
correlated albeit with rather large scatter. This is shown,
in particular, by \citet{Cross02}, based on bivariate distributions of
$M_{\rm B}$, $\mu_{B,0}$. The ridge of this bivariate distribution
along which $\mu_{B,0}$ systematically changes with $M_{\rm B}$ is read as
\begin{equation}
$$\mu_{B,0}-\mu_{B,0}^{*} = \beta_{\mu} \times (M_{\rm B}-M_{\rm B}^{*})$$,
\label{eq:ridge}
\end{equation}
where $\beta_{\mu}$=0.281$\pm$0.007, $\mu_{B,0}^{*}$=22.65~mag~arcsec$^{-2}$
and $M_{\rm B}^{*}$=--20.20. The r.m.s scatter in $\mu_{B,0}$ around this
ridge is estimated as $\sigma_{\mu}$=0.517$\pm$0.006~~mag~arcsec$^{-2}$.
All the values above were transformed
to $B$-band from the original $b_{J}$ magnitudes of the 2dFGRS (Two-degree
Field Galaxy Redshift Survey) in
\citet{Cross02} using the transform of $B \approx$ $b_{J}$ + 0.20~mag.
The latter is based on the relation $B = b_{J}+0.28(B-V)$ from
\citet{Hewett95} and the typical late-type galaxy colour $(B-V) \sim$ 0.7.
Using these results, we can address the possible incompleteness of the
current Lynx-Cancer void galaxy sample.

\citet{Blanton05} and \citet{Geha06} indicated that the SDSS spectral
survey of galaxies is significantly incomplete for the average SB within
$r$-filter half-light radius dimmer than 23.5~mag~arcsec$^{-2}$.
We use the effective surface brightness $\mu_{\rm eff}$
to derive an  estimate of the central SB $\mu_{B,0}$  for comparison
with the above study. For this we adopt the colour $g-r \sim$0.5 as
a value typical of late-type dwarfs. We also use the \citet{Lupton05}
formula
% ($B = g + 0.3130\cdot(g - r) + 0.2271$;  $\sigma$ = 0.0107)
to convert the
SDSS $g$ and $r$ magnitudes to Johnson-Cousins $B$.
For the adopted $(g-r)$ colour, we get a typical value of
$B-g \sim$0.35 and respectively, $B-r \sim$0.85. From this,
$\mu_{\rm eff,r}$=23.5~mag~arcsec$^{-2}$ to first approximation
corresponds to $\mu_{\rm eff,B}$=24.35~mag~arcsec$^{-2}$.
For purely exponential disks with the scalelength $\alpha_E$, the radius of
 the disk including its half-light is equal to $r_{\rm eff}$ =
1.678$\alpha_E$ \citep[e.g.][]{Impey88}. Using the definition of
$\mu_{\rm eff}$, one can show that for  purely exponential disks
this parameter is related to the central surface brightness:
$\mu_{\rm eff}$ =$\mu_{0}$+1.12.
Thus, the above assertion of  \citet{Blanton05} and \citet{Geha06} can
be formulated as a significant incompleteness of the SDSS spectral targets
for late-type dwarfs having the visible central  SB (uncorrected for
the Galaxy extinction and inclination) $\mu_{0,B} >$23.23~mag~arcsec$^{-2}$.

\begin{figure}
 \centering
 \includegraphics[angle=-90,width=8.0cm]{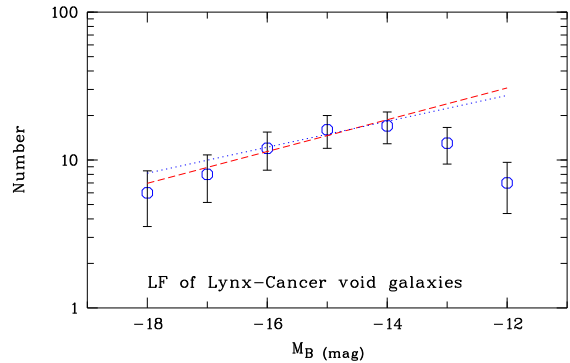}
  \caption{\label{fig:LF}
The shape of the raw $B$-band luminosity function for the Lynx-Cancer void
galaxies,
fitted by the Schechter functions on the $M_{\rm B}$ range [--14.0,--18.0].
Red dashed line with $\alpha$=--1.27$\pm$0.10 corresponds to the standard
$M_{\rm B}^{*}$=--20.2 (see text). Blue dotted line with
$\alpha$=--1.22$\pm$0.09 corresponds to the `reduced' void value of
$M_{\rm B}^{*}$=--19.2.
The large deficit of galaxies with $M_{\rm B} \gtrsim$
--13.5~mag at least partly is due to the SDSS spectral database selection
against LSB galaxies (see text).}
\end{figure}

The latter value of $\mu_{0,B}$, according to the above ridge relation
(equation \ref{eq:ridge}), corresponds to an
average galaxy with $M_{\rm B} \sim$--18.1. This implies that practically
all face-on galaxies (with small contribution of bulge emission) in the
void sample are subject to the SDSS spectroscopy SB selection.
The most probable observed inclination angles for the random orientation
sample fall around $i$=60\degr. Therefore, one expects a typical increase
of the visible surface brightness of
\mbox{$-2.5\log$($\cos~i$)$\sim$0\fm75}. The inclination brightening
(and the possible presence of central SF regions) thus should mainly
eliminate the SDSS SB selection for more luminous void galaxies. However,
it would remain crucial  for the fainter void galaxies.

Despite the fact that half of selected Lynx-Cancer void
galaxies appear to be LSB galaxies, it is quite evident
that LSB dwarfs are under-represented in this sample.
This is substantiated by the observation that no SDSS
redshifts could be obtained for several of these LSB dwarfs
but are found in the literature
as mainly measured via dedicated \HI-line observations or through
resolved stellar photometry. The examples are LSB dwarfs (LSBDs) UGC~3966
(DDO~46) with \HI\ velocity from \citet{Springob05}, SAO~0822+3545
with  H$\alpha$ velocity from \citet{Pustilnik03}, UGC~4426 with
photometric distance from \citet{ikar06}, UGC~5209 with
\HI\ velocity from \citet{HKK03}.

The effect of the SDSS spectral database magnitude selection for our void
sample can be roughly estimated using the adopted Petrosian magnitude limit
$r$=17.77.
For the great majority of galaxies with sizes less than 1\arcmin--2\arcmin,
Petrosian magnitudes do not differ more than 0.1~mag from the total galaxy
magnitudes. Therefore, using the average colour $(g-r) \sim$0.5 for
late-type galaxies, we transform with the \citet{Lupton05} formula
the limiting value $r$=17.77 to the approximate limiting value $B$=18.6.
The distances to the near and distant borders of the Lynx-Cancer void and
to its centre, $\sim$10, $\sim$26, and 18~Mpc correspond
to distance moduli of $\mu \sim$30.0, 32.0 and 31.3~mag. This implies,
that for the front void `hemisphere', the SDSS should select [if the
additional anti-selection of LSB galaxies (LSBGs) would not work] all
galaxies with $M_{\rm B} \lesssim$--12.7, and respectively, for
the whole void volume, all galaxies with \mbox{M$_{\rm B} \lesssim$--13.4}.

In Fig.~\ref{fig:LF}, we show the first approximation of the shape for the
void galaxy luminosity function (LF) based on the Table~\ref{tab:voidgal}
absolute magnitudes. This is fitted by the Schechter function
\citep{Schechter76} in the range
\mbox{--14.0 $>$ M$_{\rm B} >$--18.0}. Two values of $M_{\rm B}^{*}$ are
adopted, the standard one of \mbox{--20.2~mag} and the `reduced' one of
\mbox{--19.2~mag}, suitable for the void conditions. The LF does not look
to be affected by incompleteness down to $M_{\rm B} \sim$--14.0.
The Schechter LF low-luminosity slopes $\alpha$ for these two
values of $M_{\rm B}^{*}$ are \mbox{--1.27$\pm$0.10} and
\mbox{--1.22$\pm$0.09}, respectively.
This slope is close to that of the raw LF for the SDSS low-luminosity galaxy
sample from \citet{Blanton05}.

According to the `ridge' relation in equation~\ref{eq:ridge}, for
$M_{\rm B}=$--14.0 the average value of $\mu_{B,0}$ for disk galaxies
corresponds
to $\sim$24.4~mag~arcsec$^{-2}$, that is more than 1~mag~arcsec$^{-2}$ dimmer
than the derived estimate for the SDSS central surface brightness `threshold'.
Probably, the significant effect of visible brightening due to galaxy
inclination and the presence of central star-forming regions
counteracted the SDSS selectivity against LSBGs.
However, for galaxies with \mbox{--12.0 $> M_{\rm B} \gtrsim$--13.5},
our sample could be missing up to 25--30 objects according to the
extrapolation of the LF to the lower luminosity range.

There are at least three possible means to find `numerous' missed
generic LSB dwarfs within the boundaries of this void.
They include: (1) the blind \HI\ surveys  (e.g. those mentioned in
Section~\ref{sec:nearbyvoids}); (2a) the optical spectroscopy and H$\alpha$
detection of candidate LSB galaxies;
and (2b) \HI\ pointing sensitive observations of candidate LSB galaxies.
The latter are separated  in this sky region from the SDSS image
database. Unfortunately, no reliable criteria exist to identify good
candidates in the {\it nearby} LSBDs. Therefore, both methods
(2a) and (2b) are not expected to result
in a 'high' detection rate for galaxies residing in the nearby voids.
However, all new redshifts of LSBDs will advance our understanding of
the dwarf galaxy census in the local Universe.

\subsection{Other properties of, and the prospects of studying, the void sample galaxies}

The whole set of the studied sample galaxies includes  75 dwarfs.
Observations and the subsequent analysis of this full dataset and comparison
with similar data on the control dwarf galaxy sample in a denser
environment will require a significant effort during coming years.
However, the data collected so far already suggest differences in the
evolutionary status of some Lynx-Cancer void dwarf galaxies and those
in denser surroundings. While selection effects might play a role,
it is unlikely to influence the resulting parameters of the discovered
unevolved void dwarf galaxies discussed below.

To examine possible differences in the evolutionary status of void galaxies,
we study the following parameters. The first is the gas metallicity as
traced by O/H in H{\sc ii} regions around the sites of star formation.
The second parameter, the gas mass-fraction, defined as
the ratio $M(gas)/M_{\rm bary}$, where
$M_{\rm bary}$=$M$(\HI+He)+$M_{\rm star}$.
The contribution of molecular gas in the total gas mass of dwarf galaxies
is small. While the parameter $M$(\HI+He) is well determined directly from
the \HI\ 21-cm line flux and the adopted ratio $n(He)/n(HI)$, the
parameter $M_{\rm star}$ is a more
model-dependent. However,  with good surface photometry in several
filters, like that extracted from the SDSS images and the adopted stellar
metallicity, it can be well estimated within reasonable assumptions
using popular models like PEGASE \citep{pegase2}. The third
parameter related to the evolutionary status of a galaxy is the age of
its the oldest resolved or also unresolved visible stars.
The former is hard to obtain and would require long observations even with
the Hubble Space Telescope or similar telescopes, even for the fairly
local Lynx-Cancer void galaxies.  The ground-based surface photometry
can be a reasonable alternative.
However, the age estimate require well-resolved multicolour deep galaxy
images as well as accounting for the potential contamination of nebular
emission (if present) and corrections for the dust extinction. Again, the
SDSS images of well resolved galaxies and the PEGASE package models are
quite often
suitable for estimates of this parameter. Examples of such an analysis for
two very metal-poor Lynx-Cancer void galaxies, DDO~68 and SDSS J0926+3343,
are published by \citet{DDO68_sdss,J0926}.

It is worth mentioning that a significant fraction of the void galaxies are
paired with typical projected distances in pairs of several tens of kpc.
These include the following Lynx-Cancer void list entries: (13,14), (16,17),
(27,28), (40,41), (46,47), (66,67), (77,78). Other galaxies look like they
are forming unbound but probably coherent elongated structures
(`filaments') with  total length of $\sim$1--2~Mpc. Examples are the chains
of entries: (13,14,15,24), (53,60,63,66,67), (59,70,74). The majority of the
void galaxies at the current level of its census appear, however,
quite isolated.

As mentioned in Section~\ref{sec:nearbyvoids}, several unusual and/or very
metal-poor dwarf galaxies fall in the Lynx-Cancer void volume.
These include a very low-metallicity BCG HS 0822+3542 [12+$\log$(O/H)=7.44]
and its companion, very blue (and presumably relatively young) LSB dwarf
SAO 0822+3545 \citep{Pustilnik03}. The two most metal-poor dwarf galaxies in
the Local Volume and adjacent regions, DDO~68 and SDSS J0926+3343, with the
parameters 12+$\log$(O/H) of 7.14 and 7.12, respectively, are
representatives
of this void galaxy population. Moreover, they are situated only at 1.6 Mpc
from each other. Both of them are also unusual in the blue colours of their
outer parts, indicating ages of the oldest visible stellar population of
$\lesssim$1--3~Gyr \citep{DDO68_sdss,J0926}, which is in drastic contrast
with the colours and ages of the absolute majority of other dwarf galaxies.
At least  three other Lynx-Cancer void dwarf galaxies have
12+$\log$(O/H) $\lesssim$7.3.
They include SDSS J0812+4836 \citep{IT07} and SDSS J0737+4724, SDSS
J0852+1350, and one more blue 'young' dwarf SDSS J0723+3622
(Pustilnik et al., MNRAS, submitted).

The Lynx-Cancer void interior represents a small fraction ($<$0.05) of
the whole volume to the distance $D=$26~Mpc.
The existence in this void of the concentration of extremely metal-poor
galaxies and objects that lack visible old stellar populations
is surprising. The former fact is indicative of the
physical relationship between the evolutionary status of low-mass late-type
galaxies and their global environment. In turn, the latter
fact  hints  at the possibility of retarded galaxy
formation in the void environment.

\section{Summary}
\label{sec:summ}

\begin{enumerate}
\item
The nearby Lynx-Cancer void is described as a low-density region bordering
the Local Volume at negative supergalactic $Z$ coordinates, with the overall
extent of more than 16~Mpc and distance to the void centre of 18~Mpc.
A smaller subvoid is identified with the centre position on distance of
14.6~Mpc and radius of 6.0~Mpc
\citep[close to that described in][]{Pustilnik03}.
\item
A sample of 75 late-type dwarf and four subluminous galaxies residing in
the void region is presented.
Their main observational parameters available in the literature are
collected for further study and comparison.
\item
The shape of the Lynx-Cancer void galaxy raw LF indicates a significant drop
in the galaxy number at absolute magnitudes $M_{\rm B} \gtrsim$--14. This is
probably related to the SDSS galaxy spectral database selectivity against
LSB galaxies. To remedy this drawback, we mention three possible methods.
\item
Properties of several unusual dwarf galaxies found to date in the
Lynx-Cancer void are briefly described, with the emphasis on their very low
metallicities and relatively small ages of the oldest visible stellar
populations. These data, despite being rather limited, suggest that the
void-type environment might effectively slow down the rate of evolution
of a fraction of low-mass galaxies and postpone their formation epoch.
\end{enumerate}

\section*{Acknowledgements}

We thank the anonymous referee  for suggestions and criticism
which helped to significantly improve the quality and clarity of the data
presentation and discussion. We are thankful to  D.I.~Makarov for providing
the list of groups prior publication and A.Y.~Kniazev for useful comments
on the manuscript. This work was supported through the RFBR grants No.
06-02-16617 to SAP and ALT and No. 10-02-92650 to SAP.  SAP is also
grateful for the support through
the Russian Federal Agency of Education grant No. 2.1.1/1937.
We acknowledge the spectral and photometric data used for this
study
and the related information available in the SDSS database.
The Sloan Digital Sky Survey (SDSS) is a joint project of the University of
Chicago, Fermilab, the Institute for Advanced Study, the Japan Participation
Group, the Johns Hopkins University, the Max-Planck-Institute for Astronomy
(MPIA), the Max-Planck-Institute for Astrophysics (MPA), New Mexico State
University, Princeton University, the United States Naval Observatory, and
the University of Washington. Apache Point Observatory, site of the SDSS
telescopes, is operated by the Astrophysical Research Consortium (ARC).
This research has made use of the NASA/IPAC Extragalactic
Database (NED), which is operated by the Jet Propulsion Laboratory,
California Institute of Technology, under contract with the National
Aeronautics and Space Administration. We also acknowledge the usage of
the HyperLeda database (http://leda.univ-lyon1.fr).

\input LCvoid_tab2

\appendix

\section[]{Lists of galaxies delineating the Lynx-Cancer void}

\label{App1}

In Tables \ref{tab:lum} and \ref{tab:groups}, we summarize the
parameters of objects delineating the Lynx-Cancer void, which include,
respectively, 10 `isolated' luminous galaxies and 34 pairs  and groups
 hosting of luminous galaxies.
The following information is presented in columns of Table \ref{tab:lum}.
Column 1 - the galaxy name. Columns 2 and 3 - J2000 R.A. and Declination,
respectively. Columns 4 and 5 - heliocentric velocity and its error (as
extracted from NED). Columns 6 and 7 - the velocity relative to the LG
centre and the  adopted `distance' velocity (D(Mpc)=V$_{\rm dist}$/73).
Columns 8 and 9 - the total blue magnitude and the adopted $B$-band the
Galaxy extinction correction for the absolute blue magnitude in column 10.
In column 11 the `NED based' galaxy morphological type is shown.
In column 12, the distance in Mpc to the nearest luminous galaxy is
presented.
In column 13,  some alternative galaxy names are given. Three galaxies
with formal M$_{\rm B} >$ --19.0 are included, since the internal
extinction correction of these  almost edge-on galaxies,
will  brighten them above the adopted threshold
absolute magnitude.

In columns of Table \ref{tab:groups}
the following information is presented.
% in self-explained columns.
Column 1 - the group name, which is either the Tully group (TG)  number
\citep[for those from]
[]{Tully08}, or the name of the brightest member for pairs/triplets/groups
from \citet{KM08,Makarov09} and \citet{Makarov10}. Columns 2 and 3 -
J2000 coordinates of group centre. Column 4 - the group heliocentric
velocity; column 5 - V$_{\rm LG}$, the group velocity  relative to
the Local Group centre. Column
6 -  the group distance velocity V$_{\rm dist}$, which are taken either
from the non-redshift measurements for \citet{Tully08} sample objects
or from V$_{\rm LG}$ with the account for the discussed in
Section~\ref{sec:borders}  V$_{\rm pec}$.

\begin{figure*}
 \centering
 \includegraphics[angle=-90,width=14.5cm]{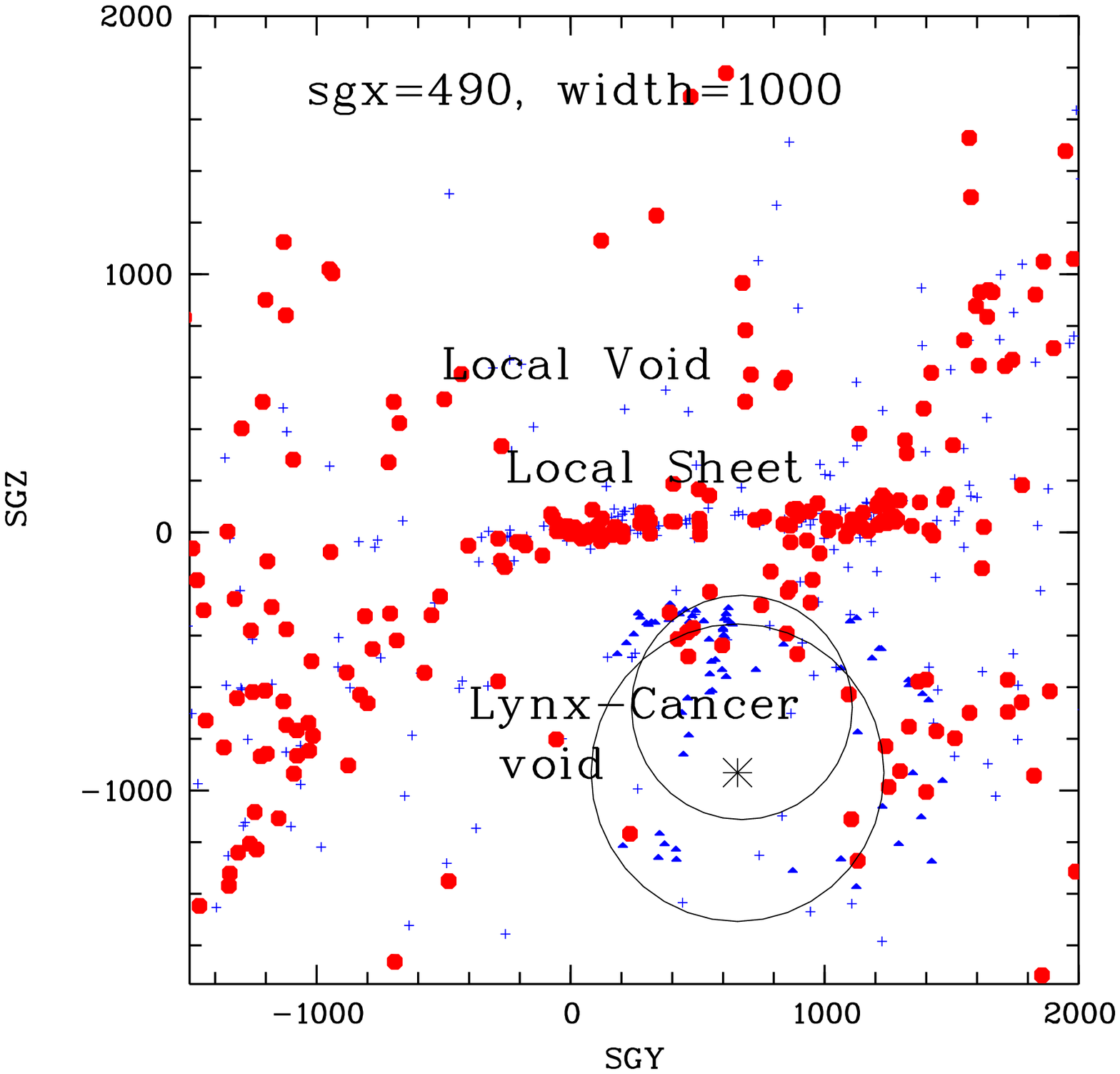}
 \includegraphics[angle=-90,width=14.5cm]{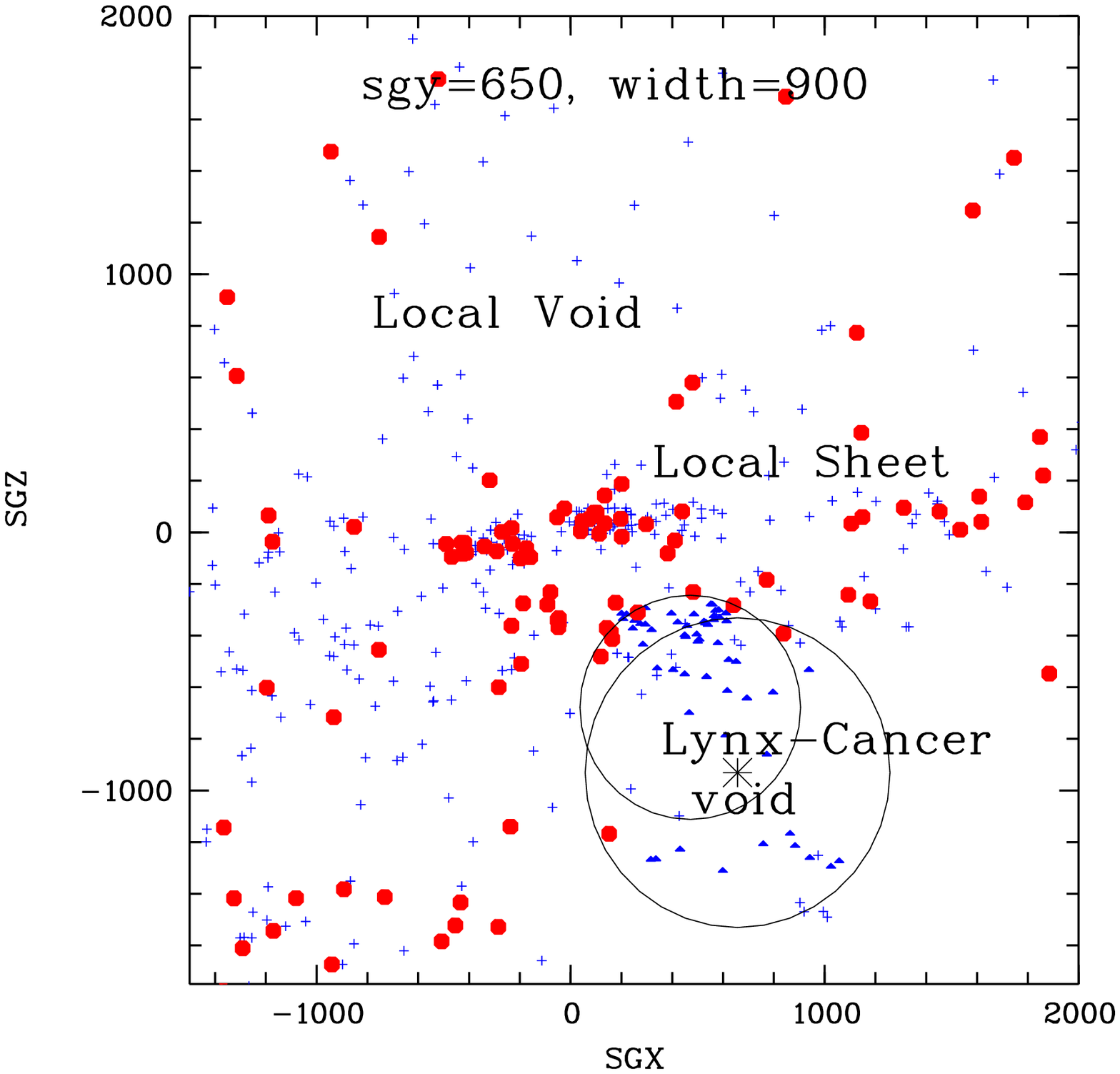}
  \caption{\label{fig:slices} Illustration of the Lynx-Cancer void position
relative to the more common elements of the local large-scale structure.
{\it top panel:} the SGY-SGZ (in \kms) projection of the slice with
the centre at SGX=490~\kms\ and the width $\Delta$SGX=1000~\kms.
{\it bottom panel:} the SGX-SGZ (in \kms) projection of the slice with
the centre at
SGY=650~\kms\ and the width $\Delta$SGY=900~\kms. The large red octagons
represent
luminous (M$_{\rm B} < $ --19.0) galaxies. The blue small crosses represent
fainter galaxies with the known non-redshift distances from \citet{Tully08}.
The blue filled triangles represent (not all) galaxies falling in the
Lynx-Cancer
void as described in Section~\ref{sec:void_dwarfs}. The projections of the
maximal inscribed spheres for Lynx-Cancer void (main) and the subvoid are
shown by the large and small circles. The asterisk marks the centre of the
main Lynx-Cancer void. The Inner Local Void of \citet{Tully08} at the positive
SGZ is almost adjacent to the Local Sheet at SGZ $\approx$0.
}
\end{figure*}

\input LCvoid_tabA1

\input LCvoid_tabA2

\label{lastpage}

\end{document}

%% file: LCvoid_tab2.tex
%  Version with removed information on IC2233 as probable luminous galaxy
%  and corrected parameters of IC2233 and NGC 2537 after Matthews & Uson 2008,
%  AJ, 135, 291
%
% Revised Vh and Bt version with added data sources. Need to change all
% related derived parameters, including histo_MB and LF
% TABLE 2. Main parameters of dwarf galaxies inside the Lynx-Cancer vois
% revised version with references to B-mags from literature (apart SDSS
% transformed).
% version of 30.11.2010 with major update (a dozen of galaxies changed)
% version of 09.01.2011 - void sample update: 11 dwarfs - removed due to wrong
%    velocities given in SDSS (V=1438, 1496, 1743), may be V=1349 - check, (
% since  V_dis~1700, moderate!)  V=1594, pure abs.
\clearpage
\begin{landscape}
\headsep 7cm
\renewcommand{\baselinestretch}{0.8}

\setcounter{qub}{0}

\begin{table*}
                                                               
\begin{center}

\caption{\label{tab:voidgal} Main parameters of dwarf galaxies inside
the Lynx-Cancer void}

\vspace{-0.3cm}

\footnotesize{\bf {
\begin{tabular}{rlccrrrrllllrl} \hline \\[-0.2cm]
\multicolumn{1}{c}{\#}               &                         
\multicolumn{1}{c}{Name or prefix}   &
\multicolumn{2}{c}{Coordinates (J2000)} &
%\multicolumn{1}{c}{$\alpha\,(2000)$} &
%\multicolumn{1}{c}{$\delta\,(2000)$} &
\multicolumn{1}{c}{$V_{\rm hel}$}  &
\multicolumn{1}{c}{$\sigma_{\rm V}$}  &
\multicolumn{1}{c}{$V_{\rm LG}$}  &
\multicolumn{1}{c}{$V_{\rm dist}$}  &
\multicolumn{1}{c}{$B_{\rm tot}$}  &
\multicolumn{1}{c}{$A_{\rm B}$}  &
\multicolumn{1}{c}{$M_{\rm B}$}  &
\multicolumn{1}{r}{Type}       &
\multicolumn{1}{r}{$D_{\rm NN}$} &
\multicolumn{1}{l}{Other names and notes}       \\
&
\multicolumn{1}{c}{ (1) }  &                                   
\multicolumn{1}{c}{ (2) }  &                                   
\multicolumn{1}{c}{ (3) }  &                                   
\multicolumn{1}{c}{ (4) }  &                                   
\multicolumn{1}{c}{ (5) }  &
\multicolumn{1}{c}{ (6) }  &
\multicolumn{1}{r}{ (7) }  &
\multicolumn{1}{r}{ (8) }  &
\multicolumn{1}{r}{ (9) }  &
\multicolumn{1}{c}{ (10)}  &
\multicolumn{1}{c}{ (11)}  &
\multicolumn{1}{c}{ (12)}  &
\multicolumn{1}{c}{ (13)} \\
\\[-0.2cm] \hline \\[-0.2cm]
%#    Name or          J2000.0 coordinates         Vhel  sigV   VLG    Vdis   B_t   A_B     M_B    Ty      D_NN     Notes
%     prefix
%      1                  2           3             4   5     6      7      8     9      10     11      12
%=====================================================================================================================================================
\qq& PGC2807187     & 06 21 03.51&  +20 10 16.6& 1318&  7&  1263&  1513& 18.00$^b$ & 3.63& -17.21& Scd   &12.76 &  ADBS J062103+2010 \\ % (NED, on ADBS V_h=2292), but 1338 on HIPASS (Kar. etal 2008)  Scd
\qq& HIPASSJ0626+24 & 06 26 20.97&  +24 39 20.0& 1473&  7&  1432&  1682& 17.60$^b$& 1.79&  -16.0 & Scd   &11.39 &                    \\ %
\qq& PGC1689759     & 06 29 58.23&  +23 34 28.5& 1452&  6&  1426&  1680& 17.10$^b$& 1.18&  -15.89& Scd   &11.36 &                    \\ %
\qq&  UGC3475       & 06 30 28.86&  +39 30 13.6&  487&  1&   524&   752& 14.97$^c$& 0.79&  -15.88& Sm    & 5.11 & MCG~7-14-2  \\ %  MHI/L_B ~2.0, O/H=7.84+-0.06, BTA
\qq&  UGC3476       & 06 30 29.22&  +33 18 07.2&  469&  4&   477&   718& 14.96$^d$& 1.02&  -16.02& Im    & 4.89 & CGCG 175-2 \\ % D(bs)=7 Mpc  MHI/L_B ~1.3  there are good HII regions!
\qq&  UGC3503       & 06 38 01.40&  +22 39 06.0& 1389&  4&  1347&  1608& 15.10$^e$& 0.66&  -17.27& Sd    &10.96 &            \\ %  Bt=15.10 (NED)                                              Sdm: 1.7x0.4'
\qq&  UGC3501       & 06 38 38.40&  +49 15 30.0&  449&  6&   528&   735& 17.20$^f$& 0.50&  -13.31& Im    & 4.24 & MCG~8-12-031 \\ %O/H=7.56+-0.07 (semi) or 7.53+-0.29 BTA 26.11.08  V_dist=818 => D=11.2 Mpc = mu=30.25
\qq&  UGC3516       & 06 43 08.51&  +22 52 24.9& 1287&  7&  1226&  1491& 16.97$^c$& 1.19&  -15.79& Sd    &10.78 &            \\ %  Bt=17 (NED) Sdm: 1.3x0.5' but V_LG=1312 (LEDA)
\qq&  KKH 38        & 06 47 54.88&  +47 30 50.0&  451&  2&   518&   725& 17.40$^g$& 0.40&  -12.99& Ir    & 4.31 & LEDA 2807122  \\ %
\qq&  UGC3587       & 06 53 54.70&  +19 17 59.0& 1267&  1&  1194&  1470& 13.84$^e$& 0.40&  -18.08& S?    &10.47 &            \\ %  Springbo et al 2005 FHI=49.57+-2.24  D(TF)=24 Mpc and V_dist=1752 - looks too large! D=20 (dm=31.5, after dV=-290 is more trustable. M_B - this case)
\qq&  UGC3600       & 06 55 40.00&  +39 05 42.8&  412&  4&   435&   679& 16.18$^h$& 0.29&  -13.95& Im    & 3.85 & PGC 019871 \\ % D(bs)=7.3 Mpc MHI/L_B ~1.7
\qq&  UGC3672       & 07 06 27.56&  +30 19 19.4&  994&  4&   968&  1236& 15.40$^i$& 0.32&  -16.08& Im    & 8.54 & PGC 20154  \\ % BTA 031224  O/H=7.89+-0.11  MHI/L_B ~1.2: Coord. of 2 HII-regs in vZee_etal 1997, AJ_114_2479, Bt and MH/L_B from vZee 2000, AJ 119, 2757. O/H=8.01+-.04 and 8.02+-.10
\qq&  UGC3698       & 07 09 16.80&  +44 22 48.0&  422&  1&   465&   701& 15.41$^h$& 0.42&  -14.92& Im    & 3.82 &  Pair with NGC 2337 \\ % D(bs)=7.2 Mpc Border of void  FHI (1992ApJS...81...5S)
\qq&  NGC2337       & 07 10 13.60&  +44 27 25.0&  436&  1&   479&   715& 13.48$^h$& 0.38&  -16.85& IBm   & 3.91 &  Pair with UGC 3698 \\ % D(bs)=7.9 Mpc Border of void
\qq&  UGC3817       & 07 22 44.48&  +45 06 30.7&  437&  1&   478&   717& 15.96$^h$& 0.44&  -14.44& Im    & 3.71 & PGC 20852 \\ %  LSBD D(bs)=8.6 Mpc. dm=29.66  MHI/L_B~2.5 (Huchtmeier_etal_03) O/H=7.75+-0.10
\qq&  SDSS          & 07 23 01.42&  +36 21 17.1&  885&  3&   882&  1050& 17.01$^q$& 0.23&  -14.19& Sm?   & 6.55 &                \\
\qq&  SDSS          & 07 23 13.46&  +36 22 13.0&  974&  3&   971&  1050& 19.25$^q$& 0.23&  -11.95& dI    & 6.55 & Companion? of J0723+3621 \\
\qq& PGC020981      & 07 25 38.95&  +09 10 59.8& 1202&  6&  1064&  1363& 16.69$^c$& 0.27&  -14.94& I     & 8.18 &  CGCG 057-013      \\ %  bt=15.7 (NED) but V_LG=1124 (LEDA)      V_rot=161; Bt now from LEDA (Paturel etal. 2000)
\qq&  UGC3853       & 07 27 39.26&  +48 26 45.4&  936&  4&   992&  1231& 15.96$^j$& 0.46&  -15.65& Sdm   & 4.20 & MCG~8-14-018=RFGC 1217         \\ %  Bt=15.83  Image of Galaxies data-base II. (Garnier+ 1996) AAS, 117, 467
\qq&  UGC3860       & 07 28 17.20&  +40 46 13.0&  354&  1&   371&570$*$& 14.96$^e$& 0.25&  -14.75& Im    & 2.51 & DDO~43  \\ %  =SDSS J072817.72+404611.3    D(rgb)=7.81 Mpc (CNG)
\qq&  UGC3876       & 07 29 17.49&  +27 54 01.9&  854&  6&   806&  1096& 13.70$^e$& 0.19&  -17.39& SAd   & 7.81 & MCG~5-18-15=KARA~193 \\ % B_tot (MW and extern. extin. corr) =13.12; DNN (wrong) = 11.68; HI - aver. of 3 meas. MHI/LB=1.13  D(TF)=24 Mpc and V_dist=1752 - looks too large! for dV=-290, D=15.07 and dm=30.89
\qq&  SDSS          & 07 30 58.90&  +41 09 59.8&  874&  3&   892&  1146& 16.67$^k$& 0.27&  -14.58& dI    & 5.25 & SDSS J073058.90+410959.8 \\ %  O/H=7.95+/-0.08 on SDSS spectrum, NRT HI detection. After V_pec D=16.19, mu=31.05 Nearest "lum" gal. is J07122870+4710004 (NGC 2344), B=12.8, M_B=-18.5, with Vh=974, at ~7deg (~2.5 Mpc). total: g=16.37 r=15.99
\qq&  UGC3912       & 07 34 12.63&  +04 32 47.1& 1240&  5&  1065&  1368& 14.72$^c$& 0.23&  -16.87& S?    & 7.62 &                    \\ %  D_TF=17.1, mu=31.16 but V_LG=1134 (LEDA) IBm:  1.8x1.1'  Bt=13.3 (NED) V_rot=117
\qq&  SDSS          & 07 37 28.47&  +47 24 32.8&  476&  2&   524&   761& 18.06$^q$& 0.47&  -12.50& LSB   & 3.22 & SDSS J073728.49+472432.8  \\   % +-54 km/s (SDSS). V(BTA)=459+-21 - check! Need HI  V_dist=727, D=10  total g=17.91 r=17.56
\qq&  UGC3966       & 07 41 26.00&  +40 06 44.0&  361&  5&   370&   631& 14.44$^c$& 0.22&  -15.46& Im    & 2.37 & DDO~46 \\ %  D(bs)=6.8 Mpc
\qq&  SDSS          & 07 44 43.72&  +25 08 26.6&  749&  4&   680&   945& 18.35$^k$& 0.18&  -12.39& Ir    & 5.82 & SDSS J074443.72+250826.6 \\ % SDSS: low EWs, no 4363, blue 12"x10", EWHa=52, NII/Ha~0.04 CHb~0.0, Hg is OK
\qq&MCG9-13-52      & 07 46 56.36&  +51 17 42.8&  445&  2&   510&   737& 16.78$^k$& 0.27&  -13.51& Sm    & 3.02 & KKH~40; Comp. of MCG9-13-56 \\ %  SDSS J074656.36+511742.8
\qq&MCG9-13-56      & 07 47 32.10&  +51 11 29.0&  439&  3&   503&   730& 15.48$^k$& 0.30&  -14.90& Sm    & 2.93 & CGCG~262-28=KUG 0743+513 \\ % , with V_pec D=11.07, mu=30.22 NRT HI detection, Izotov & Thuan (2007) astro-ph/0704.3842 in 2 HII regions: O/H=7.75+-0.04 and 7.73+-0.02
\qq&  UGC4117       & 07 57 25.98&  +35 56 21.0&  773&  1&   754&  1031& 15.34$^i$& 0.20&  -15.61& IBm   & 5.24 & MCG~6-18-3=KUG 0754+360 \\ % BTA 020113 (need empir!), MHI/L_B~2.6(?): vZeeAJ_119.2757 10 HII-regions: obs. O/H, From g,r B=15.66  V=15.29
\qq&  UGC4148       & 08 00 23.68&  +42 11 37.0&  716&  3&   729&   989& 15.63$^c$& 0.18&  -15.21& Scd   & 3.73 & MCG~7-17-6=KUG0756+423  \\ % RFGC 1301  MHI/LB=2.17 O/H=7.86+-0.11 MHI_LB~2.2
\qq& NGC2500        & 08 01 53.30&  +50 44 15.4&  504&  1&   562&   794& 12.23$^c$& 0.17&  -18.13& SBcd  & 2.27 & UGC4165=KARA~224 \\ % quadrouple system with NGC 2537 (Sc pec), NGC 2541 (Sc) and NGC 2552 (Sd),SANDAGE, A., BEDKE, J. 1994
%\qq& SDSS           & 08 01 58.89&  +21 22 19.1& 1438& 38&  1343&  1649& 17.81$^k$& 0.22&  -14.18& dI    & 4.39 &  \\ %  cor.V=2157 km/s!    ~12" blue weak emissions         120.49540838  21.37199021  2
\qq& MCG7-17-19     & 08 09 36.10&  +41 35 40.0&  704&  1&   712&   976& 16.65$^c$& 0.22&  -14.20& Sc    & 3.76 & KUG~0806+417 \\ % BTA 030102 4363! O/H=7.85+-0.11, V_h=741+-131 kms NRT HI detection
\qq& SDSS           & 08 10 30.65&  +18 37 04.1& 1483&  3&  1371&  1683& 18.29$^k$& 0.16&  -13.68& Sm:   & 4.39 &   \\ %  ~12" blue edge-on m.b. poor?  SDSS J081030.65+183704.2   122.62772691  18.61783218  2
\qq& SDSS           & 08 12 39.53&  +48 36 45.4&  521&  5&   565&   807& 17.23$^k$& 0.22&  -13.21& dIr   & 2.50 & SDSS J081239.53+483645.4 \\ % with V_pec D=11.71, mu=30.34  NRT HI detected  O/H=7.28+-0.06 Izotov and Thuan (2007) astro-ph/0704.3842 (semi-empirical), ~25"x10"
\qq& NGC2537        & 08 13 14.73&  +45 59 26.3&  445&  1&   475&   720& 12.27$^c$& 0.23&  -17.93& SBm   & 2.95 & UGC4274, not pair of IC2233 \\ %  \citep{MU08}
\qq& IC2233         & 08 13 58.93&  +45 44 34.3&  553&  1&   572&781$*$& 13.05$^l$& 0.22&  -17.32& Sd    & 3.04 & UGC4278=RFGC 1340   \\  %
\qq& NGC2541        & 08 14 40.18&  +49 03 42.1&  548&  1&   594&876$*$& 12.25$^c$& 0.22&  -18.36& SABc  & 2.33 & UGC4284, pair with NGC2552?      \\ % SA(s)cd, LINER, D=11.22 => V_dist=819 km/s, V_pec = -225 km/s; HS 0811+4913 at 1.2' NEE, w/vel 763 km/s, SA in this gal. Our O/H - from MMT and 4m KPNO (IT2003) =~8.10 (SHOC 193a,b O/H=8.23+-0.03 and 8.02+-0.01), FHI=161.0+-8.9 Springbo et al 2005 Tully 2007, add. mu=30.49+-0.35 (TF)
%\qq& SDSS           & 08 17 21.04&  +24 57 45.6& 1438& 34&  1356&  1661& 17.29$^k$& 0.17&  -14.66& dI    & 3.12 &  \\   % cor.V=2139  in SDSS Vh=1499+-15  emis.lines sit on strong Balmer abs. need BTA Ha-image
\qq& NGC2552        & 08 19 20.14&  +50 00 25.2&  524&  1&   574&   811& 12.92$^i$& 0.20&  -17.51& SAm   & 2.24 & UGC4325  pair with NGC2541?     \\ %, D(mem)=6.7 Mpc, a member of loose group, incl. NGC 2500, NGC 2537, NGC 2541 =SHOC 197, O/H=8.28+-0.11 FHI=29.88+-0.93 Springbo 2005
%\qq& SDSS           & 08 23 41.01&  +20 41 48.5& 1496& 24&  1390&  1702& 18.57$^k$& 0.16&  -13.43& dI    & 3.29 & pair w/J082417.94+203049.4? \\ %  cor.V=2086   ~10" bluish noisy sp. only Ha   125.92088085  20.69681158  2
%\qq& SDSS           & 08 24 17.94&  +20 30 49.4& 1438& 35&  1341&  1654& 17.42$^k$& 0.17&  -14.35& dI    & 3.35 & SDSS J082417.94+203049.4    \\ % cor.V=2230 ~10" blue   faint lines          126.07478747  20.5137328   2
\qq&  KUG~0821+321  & 08 25 04.90&  +32 01 05.1&  648& 16&   601&   894& 16.10$^k$& 0.20&  -14.54& Ir    & 4.41 & SDSS J082504.94+320105.1 \\ % looks as LSBD, LEDA M_B=-12.7 No SDSS spectrum available!
\qq&  HS~0822+3542  & 08 25 55.43&  +35 32 31.9&  720&  2&   691&   985& 17.92$^m$& 0.20&  -12.93& BCG   & 4.46 & Pair with SAO~0822+3545 \\ % HI from GMRT MHI/LB=0.63 V(LG)_system=705  D=(V(LG)-V_pec)/73=995/73=13.6 Mpc
\qq&  SAO0822+3545  & 08 26 05.59&  +35 35 25.7&  740&  1&   711&   985& 17.56$^m$& 0.20&  -13.29& Im    & 4.45 & Pair with HS~0822+3542             \\ %               HI from GMRT MHI/LB=1.31 V(LG)_system=705  total; g=17.63 r=17.53
%%%%\qq& SDSS           & 08 39 48.34&  +31 40 53.8&  627&  13&  585&   880& 16.74$^k$& 0.21&  -13.88&  dI   & 4.14 &  \\
  \\[-0.25cm] \hline \\[-0.2cm]
\multicolumn{14}{l}{ $*$ V$_{\rm dist}$ calculated from photometric distance. } \\
\end{tabular}
}
}
\end{center}

\end{table*}
\begin{table*}

\newpage
\begin{center}
\vspace{-0.5cm}
%\caption{Main parameters of dwarf galaxies inside the Lynx-Cancer void. Continued}
{\bf Table 2.} Main parameters of dwarf galaxies inside the Lynx-Cancer void. Continued
\vspace{-0.0cm}
\footnotesize{\bf {
\begin{tabular}{rlccrrrrllllrl} \hline \\[-0.2cm]
\multicolumn{1}{c}{\#}               &
\multicolumn{1}{c}{Name or prefix}   &
\multicolumn{2}{c}{Coordinates (J2000)} &
%\multicolumn{1}{c}{$\alpha\,(2000)$} &
%\multicolumn{1}{c}{$\delta\,(2000)$} &
\multicolumn{1}{c}{$V_{\rm hel}$}  &
\multicolumn{1}{c}{$\sigma_{\rm V}$}  &
\multicolumn{1}{c}{$V_{\rm LG}$}  &
\multicolumn{1}{c}{$V_{\rm dist}$}  &
\multicolumn{1}{c}{B$_{\rm tot}$}  &
\multicolumn{1}{c}{$A_{\rm B}$}  &
\multicolumn{1}{c}{M$_{\rm B}$}  &
\multicolumn{1}{r}{Type}       &
\multicolumn{1}{r}{D$_{\rm NN}$} &
\multicolumn{1}{l}{Notes}       \\
&
\multicolumn{1}{c}{ (1) }  &                                   
\multicolumn{1}{c}{ (2) }  &                                   
\multicolumn{1}{c}{ (3) }  &                                   
\multicolumn{1}{c}{ (4) }  &                                   
\multicolumn{1}{c}{ (5) }  &
\multicolumn{1}{c}{ (6) }  &
\multicolumn{1}{r}{ (7) }  &
\multicolumn{1}{r}{ (8) }  &
\multicolumn{1}{r}{ (9) }  &
\multicolumn{1}{c}{ (10)}  &
\multicolumn{1}{c}{ (11)}  &
\multicolumn{1}{c}{ (12)}  &
\multicolumn{1}{c}{ (13)} \\
\\[-0.2cm] \hline \\[-0.2cm]
%#    Name or      J2000.0 coordinates         Vhel  sigV   VLG    Vdis   B_t   A_B     M_B    Ty      D_NN     Notes
%     prefix
%      1              2           3             4      5     6      7      8     9      10     11      12
%=====================================================================================================================================================
%\qq& SDSS         & 08 26 30.43&  +11 47 09.8&   1496& 16&   1345&  1665&17.27$^k$& 0.13&  -14.65& Sm:   & 6.63 &      \\ % cor.V=1965   ~16" blue face-on no emis.lines  126.62683186  11.78607465  2
\qq& UGC4426      & 08 28 28.53&  +41 51 22.8&    397&  4&    402&752$*$&15.13$^c$& 0.16&  -15.09& Im    & 2.66 & DDO~52  \\ % New D(TRGB)=10.28 Mpc (0511648_Kara_AJ_2006) FHI=10.8+-0.43 2005ApJS...160...149S
\qq&  SDSS        & 08 31 41.21&  +41 04 53.7&    582& 40&    581&   850&17.71$^k$& 0.16&  -12.78& LSB   & 3.70 & SDSS J083141.21+410453.9 \\ % NRT HI: marginal detection V(HI)=640 km/s D_LG=7.96 (NED)
%\qq& SDSS         & 08 42 37.87&  +37 35 41.0&   1438&  36&  1417&  1697&17.79$^k$& 0.15&  -14.19& Sdm?  & 3.38 &   \\ %  cor.V=2090,  ~20" blue edge-on faint lines    130.65811064  37.59519224  2
\qq&  SDSS        & 08 43 37.98&  +40 25 47.2&    614&   3&   608&   880&17.83$^k$& 0.14&  -12.72& Im    & 3.49 & SDSS J084337.98+402547.2 \\ % O/H=7.55+/-0.35 on SDSS spectrum. NRT HI marg. detection. BTA: O/H=7.60+-0.16, New O/H=7.44: Hibbard etal. 08: GBT - well detected  total: g=17.55 r=17.25
\qq&  SDSS        & 08 45 25.40&  +15 19 46.0&   1642&  50&  1504&  1824&18.60$^k$& 0.09&  -13.48&  dI   & 5.11 &   \\ %  V_meas = 1673 (Ha)
\qq& SDSS         & 08 52 33.75&  +13 50 28.3&   1511&   4&  1364&  1685&17.43$^q$& 0.16&  -14.55& Im    & 4.05 &   \\ %  ~25" blue LSB 4363! XMD?         133.14064079  13.84121887  2
\qq&  SDSS        & 08 52 40.94&  +13 51 56.9&   1541&  22&  1394&  1685&19.73$^q$& 0.16&  -12.25& dI    & 4.05 & companion of J0852+1350 \\ %  data from paper in preparation on 3 LSBDs
\qq&  UGC4704     & 08 59 00.28&  +39 12 35.7&    596&   6&   581&   857&15.51$^c$& 0.13&  -14.97& Sdm   & 3.20 & MCG~7-19-11=KUG0855+394 \\ %, RFGC 1462 very flat, size:4.1'x0.41' MH/LB=4.01, UGC: Comps:, nearest 2.1, 153, 0.5 x 0.5, V_h=473+-86, O/H=7.96+-0.17. Proj.dist to NGC2683 ~0.8 Mpc (if at the same D), but V_h is 180 km/s higher
\qq&  SDSS        & 08 59 46.93&  +39 23 05.6&    588&  34&   573&   849&16.98$^k$& 0.11&  -13.46& dI    & 3.16 & companion of UGC4704  \\ % O/H=7.57+-0.06 IT07, astro-ph/0704.3842 (semi-empirical) D(LG)=7.85+/-0.72 Mpc; (m-M)=29.48  g=17.07,r=16.77, proj.dist. to UGC 4704 ~13'=> ~30 kpc (new O/H=7.38, Hibbard etal 08: GBT - well detected! W50=39 total: g=17.00 r=16.64
\qq&  UGC4722     & 09 00 23.54&  +25 36 40.6&   1794&   7&  1728&  2036&15.16$^k$& 0.17&  -17.24& Sdm   & 4.13 &  merger? \\ %%  Nice edge-on merger with tail-plum
%%%%\qq& UGC4781      & 09 06 34.29&  +06 18 13.6&   1443&   5&  1259&  1581&15.39$^c$& 0.19&  -16.48&  Scd: & 4.67 & MCG+01-23-022=KIG~300    \\
%%%%\qq&  SDSS        & 09 08 24.01&  +06 57 05.4&   1556&   3&  1375&  1697&16.99$^k$& 0.19&  -15.03&  Sc:  & 4.83 &               \\
%\qq&  SDSS        & 09 07 06.26&  +32 22 19.5&   1743& 58&   1690&  1984&17.85$^k$& 0.09&  -14.60& dI    & 2.60 &    \\ % cor.V=2010    Ha +faint N1,N2
%\qq& SDSS         & 09 10 01.72&  +32 56 59.8&   1438&  34&  1387&  1679&17.24$^k$& 0.09&  -14.66& dI    & 2.77 &       \\ %  cor.V=2070   ~15" blue only faint lines       137.50717518  32.94994839  2
%%%%\qq& SDSS         & 09 10 28.77&  +07 11 17.9&   1523&   5&  1342&  1664&17.00$^k$& 0.19&  -14.98& dI    & 4.70 &  \\ % knot in elong.dI, strong lines, 4363! ~20x5"
\qq&  SDSS        & 09 11 59.43&  +31 35 35.9&    750&   4&   692&   987&17.97$^k$& 0.07&  -12.75& dI    & 3.90 & SDSS J091159.43+313535.9 \\ % very blue XMD, g=17.72, r=17.63, 4363/Hg~0.1 O/H=7.47+/-0.55 (SDSS) Izotov & Thuan (2007) astro-ph/0704.3842 O/H=7.51+-0.14 (plus semi-empirical O/H=7.46+-0.05)  total: g=17.72 r=17.56
%%%%\qq&  SDSS        & 09 12 50.91&  +06 28 32.9&   1469&  52&  1285&  1607&17.92$^k$& 0.21&  -14.00&  dI   & 4.19 &  \\
%%%%\qq& SDSS         & 09 14 48.80&  +33 01 15.2&   1496&  15&  1445&  1737&16.27$^k$& 0.08&  -15.69&  dI   & 2.1: &  \\
%%%%\qq&  SDSS        & 09 14 57.33&  +06 00 18.6&   1428&  35&  1242&  1564&17.66$^k$& 0.18&  -14.17&  dI   & 3.85 &  \\
\qq&  IC2450      & 09 17 05.27&  +25 25 44.9&   1644&   2&  1552&  1859&13.84$^e$& 0.14&  -18.33&  S0:  & 4.09 &  UGC~4902=MRK~1230     \\
%\qq& SDSS         & 09 20 02.66&  +28 20 57.4&   1496&  10&  1420&  1721&18.36$^k$& 0.09&  -13.59& dI    & 3.11 &   \\ % cor.V=1976 ~10" bluish faint Ha+Bal.abs     140.01110618  28.3492909   2
%%%%\qq& PGC026453    & 09 20 59.60&  +11 03 33.0&   1280&  40&  1116&  1437&15.82$^k$& 0.16&  -15.81&  Sp:  & 3.27 &   CGCG062-024 \\
%%%%\qq&  SDSS        & 09 21 14.97&  +09 43 52.2&   1400&  60&  1230&  1551&18.06$^k$& 0.20&  -13.77&  dI   & 3.72 &               \\
\qq&  SDSS        & 09 26 09.45&  +33 43 04.1&    536&   2&   488&   776&17.34$^n$& 0.08&  -12.90& LSB   & 2.89 & SDSS J092609.45+334304.1 \\ % LSB edge-on, ~30", warped. SW edge knot spectrum's very noisy, only Ha, 0.66'x0.16', HI: NRT well detected, V(HI)=536! MHI/L_B=3
%\qq& SDSS         & 09 27 34.38&  +18 26 44.8&   1438&  37&  1310&  1625&17.55$^k$& 0.15&  -14.34& dI    & 3.38 & \\ % cor.V=2497   ~10" blue E+A                    141.89328697  18.44578578  2
%\qq&  SDSS        & 09 27 55.57&  +14 45 59.4&   1438&  34&  1291&  1609&17.65$^k$& 0.14&  -14.21&  dI   & 2.73 &  \\  %  cor.V=2186    redshift disappeared in later SDSS version, then appeared again!
\qq& SDSS         & 09 28 59.06&  +28 45 28.5&   1229&  41&  1154&  1453&16.70$^k$& 0.09&  -14.88&  dI   & 2.50 &  \\  % Ha,NII,SII, Hb and N1,N2  se and PM2010 can be applied for O/H
\qq& SDSS         & 09 29 51.83&  +11 55 35.7&   1614&  51&  1454&  1773&17.36$^k$& 0.11&  -14.69&  dI   & 2.25 &  \\  %  cor.V=1614 (vacHa) ~30x15" only Ha,NII, SII   DNN = 2.25 NN_name=TG401 LargeVoid=10.91  SmallVoid= 11.90
\qq&  SDSS        & 09 31 36.15&  +27 17 46.6&   1505&   2&  1422&  1723&17.98$^k$& 0.08&  -13.96&  Sm?  & 2.34 &    \\  % 4363! 4959 << Hb, NII - not seen. Poor!?
\qq&KUG0934+277   & 09 37 47.65&  +27 33 57.7&   1594&   2&  1538&  1837&16.53$^k$& 0.08&  -15.55&  Im   & 3.23 & KKH 52, in KUG classif. as GPair   \\ % only Ha at V=1622.
\qq&  SDSS        & 09 40 03.27&  +44 59 31.7&   1246&  80&  1260&  1512&17.96$^k$& 0.06&  -13.69&  dI   & 3.99 &     \\   %  probably poor: O/H~7.30+-0.15
\qq&  KISSB23     & 09 40 12.67&  +29 35 29.3&    505&   2&   450&   745&16.32$^o$& 0.10&  -13.82& Sd    & 2.23 & KUG~0937+298 \\ % V_h from NRT HI Salzer: O/H=7.65  MHI/LB=1.05  Izotov & Thuan (2007) astro-ph/0704.3842 O/H=7.65+-0.05 after correction for V_pec=-200 km/s, D=8.93 and M_B=-13.53  total g=15.93 r=15.77
\qq&  UGC5186     & 09 42 59.10&  +33 16 00.2&    551&   1&   501&   786&16.27$^h$& 0.06&  -13.95& Im    & 2.58 & KUG~0940+334 \\ % UGC:Companions 3.7, 14, 0.4'x0.3';and 4.9, 23, 0.3'x0.3' HI-flux at 2.2 sigma
%%%%\qq&  SDSS        & 09 43 38.13&  +35 12 07.8&   1511&  20&  1471&  1752&17.81$^k$& 0.06&  -13.72& dI    & 1.99 &     \\ %  Took 2-nd value of Vh from NED (from emis.lines) D_NN=1.99 for TG401
\qq& SDSS         & 09 43 42.97&  +41 34 08.9&   1403&  40&  1398&  1662&17.63$^k$& 0.06&  -14.22& dI    & 2.11 &  \\ %    ~12" blue  no emis.              145.9290504   41.56915293  2
\qq&  SDSS        & 09 44 37.11&  +10 00 46.3&   1477&  66&  1307&  1622&16.95$^k$& 0.12&  -14.90& dI    & 2.12 &   \\  % D_NN=1.80? since V_LG=1357!  only faint Ha  V=1563
\qq&  UGC5209     & 09 45 04.20&  +32 14 18.2&    538&   1&   482&   770&16.06$^h$& 0.08&  -14.14& Im    & 2.75 & KKH~54 \\ % HI in KKH 2001
\qq& SDSS         & 09 47 18.35&  +41 38 16.4&   1389&   2&  1384&  1647&17.61$^k$& 0.07&  -14.22& dI    & 2.64 &    \\ %  ~12" blue comet-like 4363!  Kniazev result on O/H     146.82647545  41.63790921  2
\qq&  SDSS        & 09 47 58.45&  +39 05 10.1&   1479&  62&  1460&  1730&18.12$^k$& 0.07&  -13.82& dI    & 2.17 &   \\  % 1' to NEE blue gal. ~2.2mag fainter in NED Vh=1501+-60  Only abs.lines!
%%%%\qq&  SDSS        & 09 49 03.13&  +33 59 28.9&   1494&   2&  1447&  1730&17.34$^k$& 0.04&  -14.57&  dI   & 2.37 &               \\
%%%%\qq&  SDSS        & 09 49 11.29&  +34 26 34.2&   1489&   2&  1445&  1727&17.45$^k$& 0.05&  -14.47&  BCG? & 2.48 & companion of MCG+06-22-016  \\  % Dproj=5.2 arcmin = 35.7 kpc at distance 23.6 Mpc
%%%%\qq& KUG0946+346  & 09 49 36.48&  +34 26 14.3&   1488&   2&  1444&  1726&16.13$^c$& 0.04&  -15.77&  Sdm? & 2.46 & MCG+06-22-016     \\
\qq&  UGC5272b    & 09 50 19.49&  +31 27 22.3&    539&   4&   483&   750&17.68$^h$& 0.10&  -12.48& Im    & 2.51 & KK~78, companion of UGC5272 \\ % (size ~18"x10") D(mem)=7.1 Mpc,  need GMRT for HI detection/interaction. KK 78 =  SDSS J095019.49+312722.2  4363 in SDSS spectrum is tooo strong (that is noised), but can be XMD, need to check, BTA 12.01.08 4x900s total: g=17.42 r=17.23
\qq&  UGC5272     & 09 50 22.40&  +31 29 16.0&    520&   5&   464&   752&14.46$^c$& 0.10&  -15.70& Im    & 2.50 & DDO~64=MCG~5-23-41  \\ % KUG0947+317  O/H=7.86 MHI/LB=1.89  D(bs)=7.1 Mpc
\qq&  SDSS        & 09 51 41.67&  +38 42 07.3&   1435&   4&  1414&  1684&17.42$^k$& 0.08&  -14.47&  dI   & 2.62 &   \\   %  Nice spectrum, all lines are seen, need O/H
\qq& SDSS         & 09 54 50.60&  +36 20 01.9&    503&  55&   470&   746&17.93$^k$& 0.04&  -12.16&  dI   & 2.39 &  \\    %  Only Ha in emission
\qq&  UGC5340     & 09 56 45.70&  +28 49 35.0&    502&   5&   428&   720&14.60$^p$& 0.08&  -15.45& BC/Im & 2.06 & DDO~68=MCG~5-24-4=VV542 \\ %  D(bs+mem)=6.5 Mpc O/H=7.21+-0.05 (aver. on 3 knots in the ring) size:2.7'x1.0' Mak.& Kara. 1998 No galaxies distur, but morphol. is disturbed! looks like merging in blue! 2 knots in center, check kinematics; U5427 ~200 kpc in proj. 1.2 mag.fainter After correct. for V_pec=-290, Dist=9.9
\qq&  PC0956+4751 & 09 59 18.60&  +47 36 58.4&   1093&   4&  1122&  1362&17.14$^k$& 0.07&  -14.28& dI    & 2.77 & SDSS J095918.68+473658.5 \\ % =PC 0956+4751  NRT HI detected  Hg almost in abs. strong OI 6300  0.65'x0.27'
\qq&  SDSS        & 10 00 36.50&  +30 32 09.8&    501&  37&   436&   723& 18.06$^k$& 0.08&  -12.00& dI    & 2.17 & SDSS J100036.54+303209.8 \\ %  Hg too weak,small LSB EWHa=42 (SDSS); NRT HI detected, very narrow, FHI~0.12 Jy*km/s   total: g=17.74 r=17.44
\qq& KUG0959+299  & 10 02 23.18&  +29 43 33.3&    766&  45&   697&   984& 17.32$^k$& 0.10&  -13.43& dI    & 2.98 &  \\ %
\qq&  UGC5427     & 10 04 41.05&  +29 21 55.2&    498&   5&   427&   715& 14.89$^h$& 0.10&  -15.16& SABd  & 2.02 & MCG~5-24-10 \\ %  Many bright HII-knots (Kaisin) SDSS spectrum for central region: no 4363, BTA: mean on 2 HII-reg, at S-edge:7.92+-0.06
\qq&  UGC5464     & 10 08 07.70&  +29 32 34.4&   1003&   3&   948&  1234& 15.62$^e$& 0.10&  -15.62& Sm    & 3.70 & DDO~72=MCG~5-24-17 \\ % SDSS spectrum, 4363, EWHa=138
\qq&  SDSS        & 10 10 14.96&  +46 17 44.1&   1092&   3&  1114&  1356& 18.20$^k$& 0.03&  -13.17& dI    & 3.23 & SDSS J101014.96+461744.1 \\ % 0.27'x0.16'  NRT HI not detected? upper limit; O/H=7.82+/-0.07 (SDSS)
\qq&  UGC5540     & 10 16 21.70&  +37 46 48.7&   1162&   4&  1134&  1399& 14.60$^e$& 0.07&  -16.89& Sc    & 3.21 & Pair with HS~1013+3809    \\ % (Virgo infall, H=73)  SDSS spectrum (4363 - 1 sigma) O/H=8.04+-0.35
\qq&  HS 1013+3809& 10 16 24.50&  +37 54 46.0&   1173&   3&  1145&  1409& 15.99$^k$& 0.07&  -15.51& BCG   & 3.22 & KUG 1013+381, pair with UGC5540 \\ % - XMD BCG, a companion of UGC 5540 (at ~50 kpc in proj.)
\qq&  SDSS        & 10 19 28.52&  +29 23 02.3&    874&  43&   831&  1113& 17.43$^k$& 0.14&  -13.63& dI    & 2.67 &   \\  %  Ha,5007,4959. Hb in deep abs.
 \\[-0.25cm] \hline \\[-0.2cm]
%\multicolumn{14}{l}{ $^a$ absolute magnitudes are corrected for galactic extinction. } \\
\end{tabular}                                                                                                                                                                                                      
}
}
\end{center}
\end{table*}
\end{landscape}
\clearpage

%% file: LCvoid_tabA1.tex
\clearpage
\begin{landscape}
\headsep 7cm
\renewcommand{\baselinestretch}{0.8}

\setcounter{qub}{0}

\begin{table*}
                                                               
\begin{center}

\caption{\label{tab:lum} List of `isolated' luminous galaxies delineating
the Lynx-Cancer void}

\vspace{-0.3cm}

\footnotesize{\bf {
\begin{tabular}{rlllrrrrlllrrrrl} \hline \\[-0.2cm]
\multicolumn{1}{c}{\#}               &                         
\multicolumn{1}{c}{Name}   &
\multicolumn{2}{c}{Coordinates (J2000)} &
\multicolumn{1}{c}{$V_{\rm hel}$}  &
\multicolumn{1}{c}{$\sigma_{\rm V}$}  &
\multicolumn{1}{c}{$V_{\rm LG}$}  &
\multicolumn{1}{c}{$V_{\rm dis}$}  &
\multicolumn{1}{c}{$B_{\rm tot}$}  &
\multicolumn{1}{c}{$A_{\rm B}$}  &
\multicolumn{1}{c}{M$_{\rm B}^{\;\;a}$}  &
\multicolumn{1}{r}{SGXV}       &
\multicolumn{1}{r}{SGYV}       &
\multicolumn{1}{r}{SGZV}       &
\multicolumn{1}{r}{D$_{\rm NN}$} &
\multicolumn{1}{l}{Notes}       \\
&
\multicolumn{1}{c}{ (1) }  &                                   
\multicolumn{1}{c}{ (2) }  &                                   
\multicolumn{1}{c}{ (3) }  &                                   
\multicolumn{1}{c}{ (4) }  &                                   
\multicolumn{1}{c}{ (5) }  &
\multicolumn{1}{c}{ (6) }  &
\multicolumn{1}{r}{ (7) }  &
\multicolumn{1}{r}{ (8) }  &
\multicolumn{1}{r}{ (9) }  &
\multicolumn{1}{c}{ (10)}  &
\multicolumn{1}{c}{ (11)}  &
\multicolumn{1}{c}{ (12)}  &
\multicolumn{1}{c}{ (13)}  &
\multicolumn{1}{c}{ (14)}  &
\multicolumn{1}{c}{ (15)} \\
\\[-0.2cm] \hline \\[-0.2cm]
%#    Name or      J2000.0 coordinates         Vhel  sigV   VLG    Vdis   B_t   A_B     M_B          SGXV     SGYV    SGZV   D_NN   Other names, notes
%     prefix
%      1              2           3             4      5     6      7      8     9      10           11      12      13       14     15
%==============================================================================================      ======================================================================
\qq&  NGC2543    &  08 12 58.0 & +36 15 17  & 2471&    9 &2449&   1920& 12.86& 0.30 &  -19.54&      1102& 1125& -1098 & 4.90 &        \\  % mu(TF)=32.10
\qq&  NGC2654    &  08 49 11.8 & +60 13 14  & 1349&    3 &1478&   1924& 12.67& 0.28 &  -19.70&      1439& 1226&  -356 & 2.55 &        \\  % Bcorr=11.65 MBcorr=-20.15
\qq&  NGC2880    &  09 29 34.6 & +62 29 27  & 1608&   21 &1716&   1600& 12.46& 0.14 &  -19.38&      1157& 1091&  -171 & 1.59 &        \\  % V_dist=1600 D=22.0
\qq& NGC3041     &  09 53 07.2 & +16 40 40  & 1408&    2 &1271&   1829& 12.31& 0.15 &  -19.81&       208& 1476& -1060 & 2.27 &         \\  % corected to M-m=31.98
\qq& NGC3198     &  10 19 54.9 & +45 32 59  &  663&    4 & 682&   1008& 11.07& 0.05 &  -19.68&       482&  855&  -231 & 1.33 & UGC5572 \\  %size:8.5'x3.3' Tully 2007. addit. mu=30.40+-0.35 (TF) D=13.8 V_dist=1008
\qq& NGC3239     &  10 25 05.7 & +17 09 37  &  753&    3 & 622&    912& 11.71& 0.14 &  -18.91$^{*}$&  52&  798&  -438 & 2.04 & UGC5637 \\  % Bcorr=11.23 MBcorr=-19.25  ARP263, VV095, KPG236, probable merger D=12.49 mu_dist=30.48 ==> M_B_0=-18.91
\qq& NGC3319     &  10 39 09.5 & +41 41 13  &  739&    1 & 740&    971& 11.48& 0.06 &  -19.20&       387&  864&  -215 & 1.33 &          \\ % Tully 2007 addit. mu=30.92+-0.35 (TF) UGC 5789 D=13.3 V_dist=971
\qq& NGC3344     &  10 43 30.2 & +24 55 25  &  586&    4 & 498&    788& 10.38& 0.14 &  -19.93&       113&  726&  -284 & 2.40 & UGC 5840 \\ % (R)SAB(r)bc  HII almost face-on D(hubble)=6.9 Mpc  D_dist=10.8 (for dV_pec=290) mu=30.17;  mu(TF)=28.92+-0.4
\qq& NGC3365     &  10 46 12.6 & +01 48 47  &  986&    1 & 789&   1336& 13.17& 0.20 &-18.34$^{*}$&  -294& 1116&  -674 & 4.85 &          \\  % Bcorr=11.67 MB=-19.13 Tully 2007 addit. mu=30.80 (2M) UGC 5878 = RFGC 1876, M_B with corr. for inter. extin. =1.25 mag D=14.45 V_dist=1055
\qq& NGC3432     &  10 52 31.0 & +36 37 10  &  613&    4 & 606&   1117& 11.64& 0.06 &  -19.33 &      341& 1031&  -259 & 2.45 &          \\  % Bcorr=10.57 MB=-19.78  Adopt. D=15.3 and <mu>=30.91 on 3 data in  1997ApJS..109..333W
\\[-0.2cm] \hline \\[-0.2cm]
\multicolumn{15}{l}{ $^a$ absolute magnitudes are corrected only for Galactic extinction according to NED;} \\
\multicolumn{15}{l}{ $^{*}$ Almost edge-on galaxies. The account for internal extiction leads to M$_{\rm B} < $ --19.0. }
\end{tabular}
}
}
\end{center}

\end{table*}

\end{landscape}
\clearpage

%% file: LCvoid_tabA2.tex
% Version of 7.04.2010. Corrections are made after NED non-vel. distances used
% TABLE 3. Main parameters of groups and pairs delineating the Lynx-Cancer void
\clearpage
\begin{landscape}
\headsep 7cm
\renewcommand{\baselinestretch}{0.8}

\setcounter{qub}{0}

\begin{table*}
                                                               
\begin{center}

\caption{\label{tab:groups} Main parameters of groups and pairs
delineating the Lynx-Cancer void}

\vspace{-0.3cm}

\footnotesize{\bf {
\begin{tabular}{rlccrrrrrrrl} \hline \\[-0.2cm]
\multicolumn{1}{c}{\#}               &                         
\multicolumn{1}{c}{Name}   &
\multicolumn{2}{c}{Coordinates (J2000)} &
%\multicolumn{1}{c}{$\alpha\,(2000)$} &
%\multicolumn{1}{c}{$\delta\,(2000)$} &
\multicolumn{1}{c}{$V_{\rm hel}^{\,a}$}  &
\multicolumn{1}{c}{$V_{\rm LG}^{\,a}$}  &
\multicolumn{1}{c}{$V_{\rm dis}^{\,a}$}  &
\multicolumn{1}{r}{SGXV}       &
\multicolumn{1}{r}{SGYV}       &
\multicolumn{1}{r}{SGZV}       &
\multicolumn{1}{r}{D$_{\rm NN}$} &
\multicolumn{1}{l}{Notes}       \\
&
\multicolumn{1}{c}{ (1) }  &                                   
\multicolumn{1}{c}{ (2) }  &                                   
\multicolumn{1}{c}{ (3) }  &                                   
\multicolumn{1}{c}{ (4) }  &                                   
\multicolumn{1}{c}{ (5) }  &
\multicolumn{1}{c}{ (6) }  &
\multicolumn{1}{r}{ (7) }  &
\multicolumn{1}{r}{ (8) }  &
\multicolumn{1}{r}{ (9) }  &
\multicolumn{1}{c}{ (10)}  &
\multicolumn{1}{c}{ (11)}  \\
\\[-0.2cm] \hline \\[-0.2cm]
%#  Name     J2000 RA   Dec                V_hel  V_LG   V_dist    SGXV   SGYV   SGZV    DNN   Notes
%    (1)        (2)        (3)              (4)    (5)     (6)      (7)    (8)    (9)   (10)    (11)
%--------------------------------------------------------------------------------------
\qq&NGC2273$^d$& 06 50 07.4& +60 50 30&         & 1967&   2190&    1917&   928&  -509&  7.44&  NGC2273 -brightest           \\  % D(N2273)=30 Mpc V_LG=1967? Check group params!
\qq&NGC2460$^c$& 07 57 02.5& +60 22 38&         & 1558&   1723&    1388&   945&  -384&  3.62&  NGC2460 -brightest           \\
\qq&IC2267 $^d$& 08 18 03.4& +24 45 33&         & 1999&   2307&     977&  1315& -1624&  6.02&  IC2267  -brightest           \\
\qq&TG 179$^a$ & 08 18 58.3& +57 48 44&     1082& 1173&    925&     708&   549&  -231&  3.02&  NGC2549 -brightest \\  % d(SBF)=12.6 d(TF)=18.8 despite large err. the latter is more trustable
\qq&TG 579$^a$ & 08 27 07.0& +25 57 27&     2068& 1988&   1876&     790&  1130& -1272&  4.90&  NGC2592 -brightest \\
\qq&TG 282$^a$ & 08 52 41.4& +33 25 19&      420&  374&    581&     272&   401&  -320&  2.16&  NGC2683 -brightest \\
\qq&TG 292$^a$ & 08 53 24.0& +51 18 54&      707&  758&   1266&     844&   857&  -395&  2.38&  NGC2681 -brightest \\
\qq&NGC2685$^b$& 08 55 35.5& +58 44 11&         &  992&   1196&     872&   782&  -240&  2.21&  NGC2685 -brightest           \\
\qq&TG 167$^a$ & 09 07 02.5& +60 07 44&     1386& 1482&   1656&    1205&  1101&  -277&  1.59&  NGC2768 -brightest \\
\qq&TG 389$^a$ & 09 09 33.5& +33 07 25&     2028& 1978&   2221&     963&  1630& -1162&  5.28&  NGC2770 -brightest \\
\qq&NGC2775$^b$& 09 10 20.1& +07 02 17&     1350& 1169&   1491&      88&   964& -1134&  4.12&  NGC2775 -brightest           \\
\qq&NGC2798$^c$& 09 17 22.3& +41 59 56&         & 1707&   1974&    1047&  1475&  -791&  3.20&  NGC2798 -brightest           \\
\qq&NGC2820$^c$& 09 21 36.7& +64 15 23&         & 1697&   1898&    1415&  1253&  -174&  2.55&  NGC2820 -brightest           \\
\qq&NGC2844$^c$& 09 21 48.7& +40 09 14&         & 1478&   1750&     877&  1329&  -727&  2.82&  NGC2844 -brightest           \\
\qq&TG 293$^a$ & 09 22 01.5& +50 59 30&      637&  684&   1029&     642&   752&  -283&  2.29&  NGC2841 -brightest \\
\qq&NGC2859$^c$& 09 24 03.0& +34 30 51&     1679& 1636&   1854&     788&  1425&  -886&  2.82&  NGC2859 -brightest           \\
\qq&TG 271$^a$ & 09 32 09.9& +21 30 00&      554&  441&    615&     140&   474&  -365&  2.16&  NGC2903 -brightest \\
\qq&TG 168$^a$ & 09 42 35.8& +58 51 08&     1362& 1451&   1090&     738&   788&  -151&  2.21&  NGC2950 -brightest \\ % d(SBF)=14.3 d(TF)=23.3 despite large err., the latter is more trustable
\qq&TG 378$^a$ & 09 45 13.5& +32 21 53&     1461& 1403&   1634&     585&  1333&  -742&  1.36&  NGC2964 -brightest \\
\qq&TG 401$^a$ & 09 50 15.0& +12 47 43&     1437& 1281&   1717&     104&  1347& -1060&  2.27&  NGC3020 -brightest \\
\qq&TG 379$^a$ & 09 52 08.3& +29 14 11&     1499& 1427&   1628&     490&  1352&  -764&  2.13&  NGC3032 -brightest \\
\qq&TG 373$^a$ & 09 55 17.8& +04 16 15&     1335& 1128&   1498&    -119&  1122&  -985&  4.46&  NGC3044 -brightest, NGC3055 \\
\qq&TG 157$^a$ & 10 01 59.5& +55 40 04&     1126& 1199&   1413&     880&  1086&  -210&  3.85&  NGC3079 -brightest \\
\qq&NGC3118$^a$& 10 07 03.6& +33 00 43&     1362& 1312&   1592&     522&  1368&  -625&  1.96&  NGC3118 -brightest \\
%\qq&TG 378$^a$& 10 07 03.6& +33 00 43&     1461& 1410&   2029&     665&  1744&  -796&  4.84&  NGC3118 -brightest \\
\qq&NGC3184$^c$& 10 17 55.2& +41 24 33&         &  592&    892&     380&   768&  -248&  1.40&  NGC3184 -brightest           \\
\qq&NGC3227$^b$& 10 23 41.4& +19 54 48&         & 1034&   1552&     159&  1366&  -718&  1.45&  NGC3227 -brightest           \\
\qq&TG 370$^a$ & 10 29 02.5& +28 44 23&     1345& 1272&   1526&     346&  1374&  -567&  2.59&  NGC3245 -brightest \\
\qq&NGC3301$^b$& 10 36 25.3& +21 49 35&         & 1219&   1504&     163&  1364&  -613&  1.45&  NGC3301 -brightest           \\
\qq&NGC3338$^b$& 10 42 27.1& +13 57 56&         & 1123&   1833&     -46&  1640&  -816&  3.76&  NGC3338 -brightest           \\
\qq&TG 266$^a$ & 10 48 15.7& +12 33 32&      819&  670&    811&     -47&   729&  -353&  1.33&  M105 group \\
\qq&TG 364$^a$ & 10 48 29.6& +12 31 07&     1285& 1136&   1565&     -93&  1407&  -679&  2.04&  NGC3373 -brightest \\
\qq&TG 268$^a$ & 10 51 23.2& +05 51 00&     1023&  844&    825&    -139&   717&  -383&  1.33&  NGC3423 -brightest \\
\qq&TG 361     & 10 56 18.6& +17 30 53&     1104&     &   1526&       7&  1416&  -569&  2.04&          \\
\qq&TG 269$^a$ & 11 00 36.2& +28 58 08&      683&  620&   1000&     177&   946&  -272&  2.55&  NGC3486 -brightest \\
\\[-0.25cm] \hline \\[-0.2cm]
\multicolumn{12}{l}{$^a$ group coordinates are from \citet{Makarov10}
[MK11]; $^b$ `groups' are from MK11, V$_{\rm dist}$ are calculated from
V$_{\rm LG}$ and $\Delta$V$_{\rm pec}$; } \\
\multicolumn{12}{l}{$^c$ MK11 groups, absent in Tully sample; $^d$ MK11
groups with a member(s),
from the Tully list. Their V$_{\rm dist}$ are adopted as for the respective
 galaxies.} \\
\end{tabular}
}
}
\end{center}

\end{table*}

\end{landscape}
\clearpage